\documentclass[11pt,a4paper]{article}
\usepackage{courier} 
\usepackage{mathrsfs}
\usepackage{setspace}
\usepackage{amsmath}
\usepackage{amssymb}
\usepackage{natbib}
\usepackage{graphicx}
\usepackage{subcaption}
\usepackage{varwidth}
\usepackage{multirow}
\usepackage{caption}
\usepackage{adjustbox}
\usepackage{lscape}
\usepackage{algorithm}
\usepackage{algorithmic}
\usepackage{geometry}
\usepackage{pgfplots}
\usepackage{float}
\usepackage{tikz} 
\usepackage{lmodern}
\usepackage[gen]{eurosym}
\usetikzlibrary{calc}

\allowdisplaybreaks

\usepackage{pgfplotstable}

\usetikzlibrary{positioning, fit, backgrounds, arrows.meta, calc, shapes.geometric}
\usepackage{booktabs}
\usepackage{siunitx}
\pgfplotsset{compat=1.17, width=15cm}
\newgeometry{margin=2.0cm}
\makeatletter
\NewDocumentEnvironment{equationseries}{ m }{%
  \ifcsname c@eqnseries#1\endcsname
    \expandafter\protected@edef\expandafter\eqnprefix\expandafter{\csname theeqnseries#1\endcsname}
    \setcounter{parentequation}{\value{equation}}
    \setcounter{equation}{\value{eqnserieseqn#1}}
  \else
    \refstepcounter{equation}
    \setcounter{parentequation}{\value{equation}}
    \protected@edef\eqnprefix{\theequation}
    \newcounter{eqnseries#1}\setcounter{eqnseries#1}{\value{equation}}
    \newcounter{eqnserieseqn#1}
    \setcounter{equation}{0}
  \fi
  \def\theequation{\eqnprefix.\arabic{equation}}
  \ignorespaces
}{%
  \setcounter{eqnserieseqn#1}{\value{equation}}
  \setcounter{equation}{\value{parentequation}}
  \ignorespacesafterend
}
\makeatother

\usepackage{hyperref}

\begin{document}

	\title{\textbf{Adaptive Robust Optimisation for Frequency-Constrained Market Clearing Problem}}

\date{}

\author{{\"O}zlem Mahmuto\u{g}ullar{\i}, Hakan Ergun \\
		{\small ELECTA, Department of Electrical Engineering, KU Leuven, 3000 Leuven, Belgium} \\
        {\small Etch, EnergyVille, 3600 Genk, Belgium}
	}

	\maketitle

\begin{abstract}

This paper formulates the day-ahead market clearing as a two-stage adaptive robust optimisation problem in which the first stage determines unit commitment, energy dispatch, and reserve procurement, while the second stage represents worst-case balancing actions under simultaneous uncertainty in renewable generation, load, and HVDC converter outages. Frequency security is enforced explicitly through inertia and fast frequency response constraints at the synchronous-area level, with HVDC lines allowed to provide emergency power support within their available headroom. The resulting min--max--min problem is solved by a column-and-constraint generation algorithm and applied to an eight-country North Sea test system comprising three synchronous areas, sixteen onshore bidding zones, and five offshore wind farm hubs connected through a meshed HVDC grid. We compare the Home Market (HM) and Offshore Bidding Zone (OBZ) designs on this infrastructure across different uncertainty scenarios. The results first demonstrate the value of robustness, where deterministic solutions that ignore uncertainty are severely vulnerable to worst-case realisations. Comparing the two designs, we find that OBZ's offshore zone boundaries make line congestion visible to the day-ahead market, enabling targeted reserve procurement that raises day-ahead costs but substantially reduces real-time load shedding, yielding a net reduction in total system cost that grows with the severity of localised offshore uncertainty. When both designs face the identical worst-case contingency, HM cannot distinguish an offshore converter trip from normal operation and thus fails to hedge against it, while OBZ considers the risk at the day-ahead stage. This advantage can be seasonal, scaling with the ratio of offshore wind output to HVDC line capacity and diminishing when onshore generation dominates the supply mix.

\end{abstract}

\section{Introduction}

The transition toward a carbon-neutral electricity system requires a massive expansion of renewable generation capacity. According to reports by the International Renewable Energy Agency (\cite{irena_tenfold}) and the International Energy Agency (\cite{iea_tenfold}), the installed capacity of inverter-based wind and solar generation is expected to increase nearly tenfold by 2050. In Europe, offshore wind energy is at the center of this transformation, and the North Sea region will host the substantial share of total European wind capacity, e.g., it is targeted to have 76 GW of offshore wind capacity by 2030, rising to 260 GW by 2050 (\cite{windEurope_2050}). This expansion will require extensive high-voltage direct current (HVDC) transmission infrastructure, which is increasingly recognised as a key enabler for large-scale renewable integration and future overlay grids (\cite{wang2020review}).

The developments in the large-scale renewable integration introduce some challenges for electricity market clearing. First, the stochastic nature of wind generation causes that day-ahead forecasts generally differ from real-time production, creating imbalances that must be resolved through costly redispatch and reserve activation. Second, the displacement of synchronous generators by inverter-based resources leads to a deficit of system inertia, making the power system more vulnerable to frequency deviations following contingency events such as generator trips or HVDC converter failures. Third, the development of meshed HVDC grids in regions such as the North Sea creates large single-point-of-failure contingencies, and thus an HVDC converter trip can instantaneously remove hundreds or thousands of megawatts of power injection from a synchronous area (\cite{gomila2025reducing}).

The market design for integrating offshore wind hubs into the electricity market can be one of the key choices to deal with the challenges posed by the uncertainties, contingencies, and low inertia. A central issue concerns the choice between two fundamentally different design paradigms, such as Offshore Bidding Zone (OBZ) and Home Market (HM) designs (\cite{kenis2023off}, \cite{hardy2023optimal}). In the OBZ design, offshore assets are placed in a separate bidding zone with its own market-clearing price. This approach explicitly acknowledges the presence of transmission capacity in hybrid offshore grids. In contrast, under the HM design, offshore wind farms are assigned to the bidding zone of the nearest onshore node, thereby implicitly assuming unconstrained access to the onshore transmission system. These design choices can affect several key aspects of system operation, including how many frequency restoration reserves are needed, how easily a system can access cross‑border balancing platforms, and what economic results can be faced (\cite{Elia_obz}). As Europe moves toward a meshed offshore HVDC system, the interaction between market zone configurations, balancing, and real-time system stability becomes increasingly critical.

In current European market practice, the day-ahead market is cleared using a zonal mechanism, typically with flow-based market coupling, that determines generation schedules and electricity prices based on forecasted conditions (\cite{van2016flow}, \cite{schonheit2021toward}). System security, including frequency stability, is ensured through separate procurement of reserves and the application of deterministic criteria such as the N-1 rule, which requires the system to withstand the loss of any single element. The management of forecast uncertainty is handled ex-post through intraday trading and real-time balancing. This sequential approach has two fundamental limitations: it does not account for the projected cost of balancing when making day-ahead decisions, and it cannot co-optimise energy dispatch, reserve procurement, and frequency stability requirements. Furthermore, balancing needs are currently determined based on static dimensioning incidents, which represent the potential outage of the largest power generation unit in the system. However, in the presence of a meshed offshore grid with large HVDC interconnectors, both loss of transmission and generation assets can constitute time-variant dimensioning incidents. The use of static indicators can thus over- or underestimate the stability impact depending on the operating conditions (\cite{gomila2025reducing}). An overestimation leads to excessive operational costs and high electricity prices, while an underestimation may result in insufficient reserves and large-scale supply interruptions.

This paper proposes a framework that addresses the aforementioned limitations. We formulate the day-ahead market clearing as a two-stage adaptive robust unit commitment problem in which the first stage determines commitment, energy dispatch, and reserve procurement, while the second stage represents balancing-stage redispatch and load-shedding actions. The uncertainty set captures variable load, renewable availability, and the loss of HVDC lines, with the change in active power following an HVDC trip endogenously determined by the day-ahead flow on the affected line. Frequency security is enforced explicitly through inertia and fast frequency response (FFR) constraints applied at the synchronous-area level, and the HVDC lines are allowed to provide FFR within available headroom. The resulting min--max--min problem is solved by a column-and-constraint generation algorithm.

The proposed algorithm is applied to the North Sea region, which includes three synchronous areas and offshore wind hubs connected via HVDC, to compare the HM and OBZ market designs on the same physical infrastructure. The comparison analyses the effect of the market design choice on total expected cost, reserve allocation, inertia provision, and the frequency-security cost of HVDC contingencies. To the best of our knowledge, this is the first study to examine the HM-vs-OBZ trade-off within a frequency-constrained market clearing problem under joint uncertainty in load, renewable availability, and HVDC line outages.

The contributions of this paper are threefold:
\begin{itemize}
    \item We propose a two-stage adaptive robust market clearing formulation that integrates flow-based zonal day-ahead clearing with synchronous-area frequency-stability constraints and treats HVDC line outages as endogenous contingencies whose post-fault active-power deficit is a decision-dependent function of the day-ahead schedule.
    \item We propose a tractable solution method based on column-and-constraint generation to solve the adaptive robust market clearing problem.
    \item We conduct a quantitative comparison of the HM and OBZ market designs on a North Sea case study, showing that the cost gap between the two designs is not welfare-neutral under robust optimisation, which is contrary to the deterministic result of \cite{kenis2023off}, because commitment and frequency security constraints create structurally different feasibility regions for the redispatch stage, with OBZ reducing total system cost by up to 21\% depending on the uncertainty level and seasonal conditions.
\end{itemize}

The remainder of the paper is organised as follows. Section~\ref{sec:literature} reviews the related literature. Section~\ref{sec:problem} describes the problem and presents the proposed mathematical formulation of the frequency-constrained robust market-clearing problem and solution method employed to solve it. Section~\ref{sec:case} reports the results of the case study conducted on the North Sea region, and Section~\ref{sec:conclusion} concludes the paper with key remarks and possible future research directions.

\section{Literature review} \label{sec:literature}

This work contributes to a literature that has so far developed in three separate streams, such as market clearing under uncertainty, frequency-constrained day-ahead scheduling, and the design of bidding zones for offshore wind integration. In this section, we review each stream and then identify the gap that motivates the paper.

\subsection{Market clearing under uncertainty}
The integration of variable renewable generation has motivated a large body of work on market clearing under uncertainty. The dominant paradigm has been stochastic programming, in which the day-ahead dispatch is jointly optimised with expected balancing costs across a discrete set of scenarios capturing possible realisations of wind generation, load, and other uncertain parameters. Early contributions such as \cite{morales2009economic} demonstrate that co-optimizing day-ahead energy and reserves within a stochastic framework yields significant cost reductions relative to traditional sequential market-clearing approaches. \cite{pritchard2010single} further show that explicitly modelling forecast errors in a two-stage stochastic dispatch framework leads to improved economic efficiency and revenue adequacy in expectation, highlighting the importance of coordinating day-ahead and real-time actions when intermittent resources are present. \cite{zavala2017stochastic} develop a stochastic market-clearing model that ensures pricing consistency between day-ahead and real-time markets by introducing penalty terms that bound price distortions and prove that such formulations mitigate systematic biases observed in deterministic clearing and improve incentives for market participants. These studies emphasise that accurate representation of balancing actions and uncertainty within the day-ahead problem is essential not only for cost efficiency but also for producing economically coherent prices and dispatch signals. Accordingly, in addition to the integrated planning of day-ahead and balancing actions, \cite{bjorndal2018stochastic} take into account the advantage of stochastic clearing on the zonal market design. The authors conclude that a stochastic model can provide significantly higher social welfare and lower day-ahead prices, which motivates the further analysis of alternative zonal designs.  
In this regard, \cite{risanger2025flexible} study the flexible electricity market clearing zones, in which the bidding-zone partition is reconfigured each clearing period to minimise expected total cost. These works confirm that the choice of zonal partition is the main determinant of dispatch and pricing outcomes under uncertainty, similar to the observation that underlies HM vs OBZ comparison.

Despite the conceptual appeal of stochastic programming, its practical use is hampered by two
well-known difficulties because it requires explicit probability distributions over uncertain parameters, and
the tractable scenario number is usually far smaller than the number of physically relevant realisations. Adaptive robust optimisation (ARO) has emerged as an alternative to stochastic programming that avoids the need for explicit probability distributions and instead protects the first-stage decision against the worst-case realisation within a deterministic uncertainty set (\cite{yanikouglu2019survey},\cite{sun2021robust}). 
\cite{bertsimas2012adaptive} introduce the two-stage adaptive robust unit-commitment problem with budget-constrained uncertainty sets and an outer-approximation heuristic. \cite{jiang2011robust} subsequently develop an exact solution approach based on binary expansion and Benders decomposition, and \cite{zhao2012robust} propose the column-and-constraint generation algorithm that generally converges faster than Benders-based approaches. Although the resulting field of study has expanded in various directions, it has mainly focused on the following uncertain dimensions: generation (\cite{zugno2015robust}, \cite{zuniga2022n}, \cite{bertsimas2024marginal}), contingency (\cite{street2010contingency},  \cite{jeong2025adaptive},
\cite{zuniga2022n}), and load (\cite{jeong2025adaptive}, \cite{bertsimas2024marginal}). However, none of these works consider frequency-stability constraints or the role of HVDC infrastructure in providing fast frequency.

\subsection{Frequency-constrained scheduling in hybrid AC/DC systems}
\label{subsec:lit_frequency}

As wind, solar, and other inverter‑based resources replace traditional generators, the power system loses natural inertia and immediate frequency support. In the literature, how this transition reduces
the system’s ability is addressed, and the system’s resilience to frequency disturbances is investigated.
The classical formulation of frequency-constrained unit commitment introduces explicit constraints on inertia, rate of change of frequency (RoCoF), nadir, and quasi-steady-state deviation, requiring the day-ahead schedule to remain feasible against a designated dimensioning incident. 
\cite{chang2012frequency} embed a linear frequency-based reserve constraint in a mixed integer linear unit commitment model for an isolated system, where fast-response reserve is assumed to be provided exclusively by synchronised pumped-storage hydropower units and its required volume is determined using an empirical load–frequency sensitivity index derived from historical contingency data. \cite{ahmadi2014security} derive a closed-form expression for the frequency nadir, linearise it through a piecewise-linear approximation, and embed the resulting constraint in the model with generic, rather than device-specific, reserve provision. \cite{wen2016frequency} extend this line of work by incorporating battery energy storage as a fast frequency support resource that injects power instantaneously after a contingency, where they employ linearised RoCoF and nadir constraints in the model accounting for wind power uncertainty.
\cite{puschel2020separation} consider separation events that split the AC system into multiple synchronous areas, which are often more severe than single generator outages. A common feature of these works is that they consider purely AC systems and treat the dimensioning incident as exogenous and fixed, typically the loss of the single largest synchronous generator.

Only a few studies extend frequency-constrained dispatch to hybrid AC/DC networks. \cite{zhang2019minimum} use minimum-frequency and voltage-stability constraints with Benders cuts but limit the analysis to small networks with point-to-point HVDC links. 
As HVDC infrastructure expands, however, converter-station and meshed offshore grid outages represent contingencies that can match or exceed the largest generator trip in severity (\cite{gomila2025reducing}), and whose post-fault active-power imbalance depends on the operating point rather than being a fixed parameter.
A separate line of work examines the potential for HVDC systems to actively support system frequency. \cite{tosatto2021sharing} demonstrate the value of sharing frequency containment reserves through HVDC emergency power control between Nordic synchronous areas. \cite{huang2017hvdc} compare droop control of HVDC converters and the exchange of frequency containment reserves, and \cite{kaushal2023economic} use a simplified market model with a fixed dimensioning incident to demonstrate the economic value of inertia support from HVDC. 
These works show that HVDC fast frequency reserve can improve  post-contingency frequency response and significantly reduce operational costs, but all of them consider the problem in a deterministic setting, disregarding the stochasticity of problem parameters.


\subsection{Market design for offshore wind: HM versus OBZ}
\label{subsec:lit_obz}

The third stream of literature concerns the market design for offshore wind integration. The status quo in the European day-ahead market is zonal clearing with flow-based coupling, in which transmission constraints are represented through a reduced set of zone-to-zone power transfer distribution factors (\cite{van2016flow}, \cite{schonheit2021toward}, \cite{byers2020modeling}). The placement of offshore generation within this zonal architecture has been actively debated. Under the HM design, offshore generators are administratively assigned to the bidding zone of the nearest onshore connection point and trade with the rest of the system as if they were located onshore, hiding offshore congestion from the zonal clearing process. Under the OBZ design, offshore assets form one or more dedicated bidding zones, exposing the limited transmission capacity of hybrid offshore links to the market and altering both dispatch and congestion-rent allocation. \cite{kenis2023off} provide a formal comparison of these designs under flow-based coupling, showing that OBZ aligns prices with physical scarcity, reduces redispatch costs, and reallocates rents between offshore generators and onshore consumers, but introduces additional complexity. \cite{kenis2024evaluating} examine how the HM vs OBZ choice interacts with offshore infrastructure investment. \cite{hardy2023optimal} evaluate the two designs across a range of North Sea topologies using a two-stage stochastic grid-expansion model. While \cite{aravena2016renewable} and \cite{aravena2021transmission} study how renewable generation interacts with zonal clearing more generally, \cite{poplavskaya2020integration} examine how flexibility on the day-ahead to balancing should be coordinated. All of these works, however, employ deterministic dispatch or scenario-based stochastic dispatch without frequency-stability constraints. The interaction between market-design choices and frequency security under uncertainty, particularly relevant given that OBZ designs can provide more flexible HVDC headroom for fast frequency reserve, has not been investigated.

\subsection{Research gap and positioning of this work}
\label{subsec:lit_gap}

Each of the three streams addresses an important aspect of the problem at hand, but their separation leaves a specific gap. The robust market-clearing literature typically assumes nodal or single-zone settings, without considering flow-based zonal clearing or explicit frequency dynamics. The frequency-constrained scheduling literature, despite including the recent treatment of hybrid AC/DC systems, is fully deterministic and cannot represent the risk under which uncertainty driven reserve activation can deplete available frequency-response headroom before a contingency occurs. The HM vs OBZ literature compares market designs on dispatch costs under deterministic or scenario-based stochastic conditions, but does not enforce frequency security and does not represent HVDC contingencies as endogenous worst-case events whose severity depends on the day-ahead flow. As a consequence, existing comparisons of HM and OBZ may misstate the true cost gap between the two designs whenever the cost of frequency security is non-negligible, because the security implications of an HVDC trip depend on which zone is, by construction, exposed to the contingency.

The present paper closes this gap by combining the three streams in a single tractable formulation. We model day-ahead clearing as a two-stage adaptive robust problem under flow-based zonal coupling, with an uncertainty set that simultaneously captures continuous deviations in load and renewable availability and HVDC converter outages. Frequency security is enforced explicitly through inertia and fast frequency response constraints at the synchronous-area level, with HVDC lines allowed to provide fast frequency response within their available headroom, where the post-trip active-power imbalance entering these constraints is endogenous to the first-stage decision. The HM and OBZ market designs are then compared on the North Sea region to investigate the effect of the market-design choice itself.

\section{Problem description and methodology} \label{sec:problem}

We consider the problem of clearing a zonal electricity market in a hybrid AC/DC grid under uncertainty in load, renewable generation capacity, and DC transmission line capacity, while ensuring frequency stability against line contingencies. The problem is formulated as a three-level adaptive robust optimisation problem with the following structure:
 
\begin{equation}
\label{three_level}
\min_{x^1 \in X^1} \; c_1(x^1) + \max_{u \in U} \; \min_{x^2 \in X^2(x^1,u)} \; c_2(x^2)
\end{equation}
 
\noindent where $x^1$ denotes the first-stage (day-ahead) decisions, $u$ represents the uncertain parameters, and $x^2$ denotes the second-stage (balancing) recourse decisions.
 
In the first-stage problem (day-ahead market clearing), the optimal day-ahead dispatch for the upcoming delivery period is determined. The decisions include the commitment status of conventional generators ($\alpha_g$), their power dispatch ($p_g$), reserve procurement ($r_g$), DC line flow schedules ($f_l$), and load shedding ($s_i$). The day-ahead market is cleared using a zonal mechanism with flow-based market coupling, where zonal power transfer distribution factors (ZPTDFs) approximate the physical AC power flows ($f^{ac}_l$). The objective minimises the total day-ahead cost, comprising generation costs, commitment costs, reserve procurement costs, and the cost of load shedding.
 
In the middle maximisation problem (uncertainty realisation), the worst-case realisation of the uncertain parameters within a predetermined polyhedral uncertainty set $U$ is selected. Three sources of uncertainty are considered: (i)~load ($d_i$) deviations, (ii)~renewable generation capacity ($P_g$) deviations, and (iii)~DC line capacity ($F_l$) deviations, where a complete failure ($F_l = 0$) models an HVDC converter contingency. The uncertainty sets are defined based on the structure of the budget uncertainty set proposed by \cite{bertsimas2004price}, with separate budgets ($\Gamma^D$, $\Gamma^P$, and $\Gamma^F$) limiting the number of simultaneously active deviations.
 
In the second-stage problem (recourse and frequency security), given the first-stage decisions and the realised uncertainty, recourse actions to restore power balance are determined. These actions include activation of reserves ($r_g^{act}$), upward and downward redispatch of renewable generators\footnote{Upward redispatch of a renewable generator is bounded by the margin between its realised available capacity and its day-ahead schedule. Since upward redispatch has a cost, this variable may take a nonzero value in two cases: (i) the realised renewable capacity exceeds the day-ahead forecast, and (ii) when the day-ahead zonal clearing curtails part of the generator’s available output, for instance due to zonal congestion management, leaving headroom that can be recovered in the nodal redispatch stage.} ($p_g^{up}$ and $p_g^{down}$), additional load shedding ($s_i^b$), and adjustment of DC and AC line flows ($f_l^b$ and $f_l^{b, ac}$). 
Crucially, the recourse problem also includes frequency security constraints that ensure the post-balancing operating point can survive the instantaneous power loss from any tripped HVDC line. These constraints bound the frequency deviation during the fast frequency response (FFR) phase using the energy-based linearisation, accounting for the inertia of committed synchronous generators and the emergency power injection capability of surviving HVDC links.

\subsection{Mathematical model} \label{sec:model}

This section presents the complete mathematical formulation of the proposed two-stage problem to deal with the uncertainty in renewable electricity generation, loads, and the status of DC lines. Table~\ref{tab:Sets}--\ref{tab:Variables} present the notation used throughout. Note that the lines are considered in two main sets, i.e., sets of AC and DC lines. AC lines have endpoints of AC buses. whereas DC lines have endpoints of DC buses or AC and DC buses with converter connections. Accordingly, each line $l$ is represented by an ordered pair of buses $(i,j)$, such that $l_1=i$ and $l_2=j$, where $i,j \in N^{ac}$ for an AC line and $i \in N^{ac} \cup N^{dc}$ and $j \in N^{dc}$ for a DC line.

\begin{table}[h!]
 	\begin{center}
  \begin{adjustbox}{width=0.85\linewidth}
 		\begin{tabular}{cll}
        Symbol       &    Description      \\ \hline
      $Z$     &    Set of bidding zones      \\
      $S$     & Set of synchronous areas \\
      $Z_s$ & Zones in synchronous area $s \in S$ \\
      {$Z^{owf}$}    &    {Set of offshore bidding zones (OBZ design only; $Z^{owf} = \emptyset$ under HM)}    \\
      $G$     &    Set of all generators \\ 
      $G^{con}$     &    Set of conventional (dispatchable) generators       \\
      $G^{res}$     &    Set of renewable energy source (RES) generators       \\
      $N^{ac}$     &    Set of AC buses      \\
      $N^{dc}$     &    Set of DC buses     \\
      $N^{ac}_z$     &    Set of AC buses in zone $z \in Z$     \\
      $N^{dc}_z$     &    Set of DC buses in zone $z \in Z$    \\
      $G_i$     &    Set of all generators at bus $i \in N^{ac} $ \\     
      $G^{con}_i$     &    Set of conventional generators at bus $i \in N^{ac}$     \\
      $G^{res}_i$     &    Set of renewable generators  at bus $i \in N^{ac}$       \\
      $L$     &    Set of DC lines  \\
      $L^{ac}$     &    Set of AC lines     \\
      $CL$     &    Set of critical AC lines, i.e.,  $CL \subseteq L^{ac}$     \\
      $CG$     &    Set of converter groups (DC buses with converters)    \\
      $L_b$    &    Set of DC lines connected to DC bus $b \in CG$    \\
      $L^{in}_s$     &    Set of DC lines $l \in L$ importing into synchronous area $s \in S$ \\
      $L^{out}_s$     &    Set of DC lines $l \in L$ exporting from synchronous area $s \in S$ \\
     \hline
 		\end{tabular}
   \end{adjustbox}
 		\caption{Sets}
 		\label{tab:Sets}
 	\end{center}
 \end{table}

  \begin{table}[h!]
 	\begin{center}
  \begin{adjustbox}{width=0.85\linewidth}
 		\begin{tabular}{cll}
         Symbol       &    Description      \\ \hline 
      $VOLL$     &   Value of lost load  \\
      $C_g$ & Commitment (no-load) cost of generator $g \in G^{con}$ \\
      $c_{g}$     &   Unit generation cost of generator $g \in G$  \\
      $c^r_{g}$     &   Unit reserve cost of generator $g \in G^{con}$  \\
      $c^{a}_{g}$     &   Unit reserve activation cost of generator $g \in G^{con}$  \\
      $c^{u}_{g}$     &   Unit upward generation cost of generator $g \in G^{res}$  \\
      $c^{d}_{g}$     &   Unit downward generation cost of generator $g \in G^{res}$  \\
      $H_{g}$     &  Inertia constant of generator $g \in G^{con}$   \\
      $P^{max}_{g}$     &   Generation capacity of conventional generator $g \in G^{con}$ \\
      $F^{ac}_{l}$     &  Capacity of AC line $l \in L^{ac}$  \\
      $\bar{P}_{g}$     &   Nominal capacity of renewable generator $g \in G^{res}$ \\
      $\bar{d}_{i}$     &   Nominal load of bus $i \in N^{ac}$ \\
      $\bar{F}_{l}$     &   Nominal capacity of DC line $l \in L$ \\
      $\hat{P}_{g}$     &   Maximum capacity deviation of renewable generator $g \in G^{res}$ \\
      $\hat{d}_{i}$     &   Maximum demand deviation of bus $i \in N^{ac}$ \\
      $\hat{F}_{l}$     &   Maximum capacity deviation of DC line $l \in L$ \\
      $\Gamma^F$     &   Maximum number of converter trips \\
      $\Gamma^D$     &   Maximum number of buses with deviating loads  \\
      $\Gamma^P$     &   Maximum number of generators with deviating capacities   \\
      $f_0$ & Nominal system frequency \\
      $\Delta f^{max}$ & Maximum allowable frequency deviation \\ 
      $f_{db}$ & Dead band frequency \\
      $\Delta T^{ffr}$ & FFR activation time \\ 
      $\Delta T^{dc}$ & HVDC ramp-up time \\ 
      {$M^{freq}$}    &   {Penalty coefficient for frequency constraint violation}    \\
      $B_{ij}, R_{ij}, X_{ij}$     &  Susceptance, resistance, and reactance of AC line between buses $i,j \in N^{ac}$  \\
      $GSK_{iz}$     &  Generation shift keys for bus $i \in N^{ac}$ and zone $z \in Z$   \\
       $NPTDF_{li}$     &  Nodal power transfer distribution factor for AC line $l \in L^{ac}$ and bus $i \in N^{ac}$\\
       $ZPTDF_{lz}$     &  Zonal power transfer distribution factor for AC line $l \in L^{ac}$ and bus $z \in Z$\\
     \hline
 		\end{tabular}
   \end{adjustbox}
 		\caption{Parameters}
 		\label{tab:Paramaters}
 	\end{center}
 \end{table}

 \begin{table}[h!]
 	\begin{center}
  \begin{adjustbox}{width=0.85\linewidth}
 		\begin{tabular}{cll}
        Symbol       &    Description     \\ \hline
      \underline{Day-ahead stage}       &         \\
      $\alpha_{g}$     &  1 if generator $g \in G^{con}$ is online; 0 otherwise   \\
      $x_{z}$     &  Net zonal injection for zone $z \in Z$     \\
      $p_{g}$     &  Generation of generator $g \in G$   \\
      $r_{g}$     &  Reserve procured from generator $g \in G$  \\
      $s_{i}$     & Load shedding at bus $i \in N^{ac}$ \\
      $E_{s}$     & Total inertia of synchronous area $s \in S$\\ 
      $f_{l}$     & Flow on DC line $l \in L$\\ 
      $f^{ac}_{l}$     & Flow on critical AC line $l \in CL$\\  \hline
      \underline{Uncertainty}       &         \\
      $F_l$     &   Capacity of HVDC line $l \in L$\\
      $d_i$     &   Load of bus $i \in N^{ac}$  \\
      $P_g$     &   Generation capacity of generator $g \in G^{res}$ \\
      $\gamma^F_l$     &   1 if HVDC line $l \in L$ is deviated from its nominal capacity; 0 otherwise\\
      $\gamma^{CG}_b$     &  1 if converter at DC bus $b \in CG$ has tripped; 0 otherwise\\
      $\gamma^D_i$     &   1 if bus $i \in N^{ac}$ is deviated from its nominal demand; 0 otherwise \\
      $\gamma^P_g$     &   1 if generator $g \in G^{res}$ is deviated from its nominal capacity; 0 otherwise \\ \hline
      \underline{Balancing stage}       &         \\
      $x^{b}_{z}$     & Net zonal injection of zone $z \in Z$ for balancing    \\
      $r^{act}_{g}$     &  Reserve activation from generator $g \in G^{con}$    \\ 
      $p^{up}_{g}$     &  Upward generation renewable generator $g \in G^{res}$   \\
      $p^{down}_{g}$     &  Downward generation renewable generator $g \in G^{res}$   \\
      $s^{b}_{i}$     & Load shedding at bus $i \in N^{ac}$  for balancing\\
      $f^b_{l}$     & Flow on DC line $l \in L$ for balancing\\ 
      $f^{b,ac}_{l}$     & Flow on critical AC line $l \in CL$ for balancing\\ 
      $\delta^+_{l}$    & HVDC FFR headroom for flow increase on DC line $l \in L$ \\ 
      $\delta^{-}_{l}$     & HVDC FFR headroom for flow decrease on DC line $l \in L$      \\
      {$\xi^+_s$}     &    {Frequency constraint slack for under-frequency in area $s \in S$}   \\
      {$\xi^-_s$}    &    {Frequency constraint slack for over-frequency in area $s \in S$}    \\
     \hline
 		\end{tabular}
   \end{adjustbox}
 		\caption{Decision variables}
 		\label{tab:Variables}
 	\end{center}
 \end{table}

The first-stage problem is related to the day-ahead market clearing and determines the day-ahead generation dispatch, reserve capacity procurement, and DC line schedules that minimise total cost, where the decisions are made according to the most expected outcome of uncertain parameters, e.g., nominal values. Accordingly, the problem can be formulated as follows:

\begin{equationseries}{firstStage} \label{first} 
\begin{align}
\text{ min } & \sum_{g \in G} c_gp_g + \sum_{g \in G^{con}} (C_g\alpha_g+ c^r_gr_g)  + \sum_{i \in N^{ac}} VOLLs_i  \label{first_obj}\\
\text{ s.t. } &  x_z=\sum_{i \in N_z^{ac}} \left(\sum_{g \in G_i} p_g  - \sum_{l \in L: l_1=i} f_{l} - \bar{d}_i +s_i \right)  \hspace{0.5cm} \forall  z \in Z \label{first_ACzonal}\\
&  \sum_{i \in N_z^{dc}} \left(  \sum_{l \in L: l_2=i} f_{l}- \sum_{l \in L: l_1=i} f_{l} \right) = 0  \hspace{1cm} \forall  z \in Z \label{first_DCzonal} \\
&  \sum_{z \in Z} x_z =0  \label{first_marketClearing}   \\
& {x_z = 0 \hspace{1cm} \forall z \in Z^{owf} \label{first_obz_netzero}}\\
&  f_l^{ac} = \sum_{z \in Z} ZPTDF_{lz}x_z  \hspace{1cm} \forall  l \in CL \label{first_CL}\\
& p_g + r_g  \leq \alpha_gP^{max}_g  \hspace{1cm} \forall  g \in G^{con} \label{first_capCon}\\
& p_g \leq \bar{P}_g  \hspace{1cm} \forall  g \in G^{res} \label{first_capRes}\\
& s_i \leq \bar{d}_i  \hspace{1cm} \forall  i \in N^{ac} \label{first_shedding} \\
& E_{s} = \sum_{z \in Z_s} \sum\limits_{i \in N^{ac}_z} \sum\limits_{g \in G_i^{con}} \alpha_gH_g \hspace{1cm} \forall  s \in S \label{first_inertia}\\
& -{F^{ac}_{l}} \leq f_l^{ac} \leq {F^{ac}_{l}} \hspace{1cm} \forall l \in  CL \label{first_limitCL}\\
& -\bar{F}_{l} \leq f_l\leq \bar{F}_l \hspace{1cm} \forall l \in  L \label{first_limitL}\\
& \alpha_g \in \{0,1\} \hspace{1cm} \forall g \in G \label{first_domainAlpha} \\
& x_z \text{ is free} \hspace{1cm} \forall z \in Z \label{first_domainInjection}  \\
& E_s \geq 0 \hspace{1cm} \forall s \in S \label{first_domainInertia} \\
& s_i \geq 0 \hspace{1cm} \forall i \in  N^{ac} \label{first_domainSheddin}  \\
& r_g \geq 0 \hspace{1cm} \forall g \in G^{con} \label{first_domainReserve}  \\
& p_g \geq 0 \hspace{1cm} \forall g \in G \label{first_domainGeneration} 
\end{align} 
\end{equationseries}

The objective function \eqref{first_obj} minimises the total day-ahead electricity generation, commitment, reserve procurement, and load shedding costs. Constraints \eqref{first_ACzonal} define the net power injection for each zone, where it is assumed that converters are lossless. Constraints \eqref{first_DCzonal} ensure the zonal energy balance for the DC network. Constraint \eqref{first_marketClearing} is the market clearing constraint that guarantees system energy balance. Constraints \eqref{first_obz_netzero} enforce that offshore bidding zones under OBZ have zero net position. Since these zones contain only offshore wind generators with no AC connections to shore, all power must be evacuated through the DC lines explicitly modelled in the zonal balance. Under HM, these constraints are inactive because offshore generators are assigned to onshore zones where their output is aggregated with onshore generation and can flow through AC lines without explicit DC capacity limits. Constraints \eqref{first_CL} determine the flows on the critical AC lines. Constraints \eqref{first_capCon} and \eqref{first_capRes} set generator capacities for conventional and renewable generators, respectively, where the conventional generator capacities are respected by reserve procurements as well. Constraints \eqref{first_shedding} set nominal demands as upper bounds for demand sheddings. Constraints \eqref{first_inertia} determine the total inertia per synchronous zone. Constraints \eqref{first_limitCL} and \eqref{first_limitL} set limits to the flows on AC and DC lines, respectively. Note that, although the constraints are given for the same limit for both directions of the 
lines, different limits can be imposed for opposite directions. Constraints \eqref{first_domainAlpha}-\eqref{first_domainGeneration} are domain constraints.

We incorporate uncertainty in RES generation, load, and converter failure into the problem through three polyhedral uncertainty sets, each following the budget-of-uncertainty structure proposed by
\cite{bertsimas2004price}. In this framework, each uncertain parameter can deviate from its nominal value by at most a specified maximum amount, and a budget parameter $\Gamma$ limits how many parameters can deviate simultaneously.  The load uncertainty set \eqref{UncertaintyLoad} implies that each bus load $d_i$ may increase above its forecast $\bar{d}_i$ by up to $\hat{d}_i$, with at most $\Gamma^D$ buses deviating simultaneously. Only upward deviations are considered because higher-than-expected demand is the cost-increasing direction. The RES uncertainty set \eqref{UncertaintyRES} implies that each renewable generator's available capacity $P_g$ may decrease below its forecast $\bar{P}_g$ by up to $\hat{P}_g$, with at most $\Gamma^P$ generators deviating simultaneously. Only downward deviations are considered because lower-than-expected renewable output forces the system to activate more expensive conventional generation. 
The line capacity uncertainty set \eqref{UncertaintyLine} models HVDC converter outages and their cascading effects on DC line availability in a meshed network. When a converter trips, it isolates the DC bus to which it is connected, and all DC lines attached to that bus lose their capacity simultaneously. To capture this physical coupling, we group DC lines by the DC bus of their shared converter. Let $CG$ denote the set of such converter groups, and let $L_b$ denote the set of DC lines connected to DC bus $b \in CG$. A binary variable $\gamma^{CG}_b$ indicates whether the converter at bus $b$ has tripped. 
The inequality constraint \eqref{lineConv} ensures that 
if converter $b$ trips, every line $l \in L_b$ is also marked as failed. 
Note that a line connecting two DC buses belongs to two converter groups. Hence, 
the inequality, rather than equality, correctly models that the line fails 
if either converter trips, without forcing both converters to share 
the same status.
The budget $\Gamma^F$ limits the number of simultaneous converter trips, directly corresponding to an N-$k$ security criterion for HVDC converters. Setting $\hat{F}_l = \bar{F}_l$ ensures that a tripped line loses its entire capacity.

\begin{equationseries}{UncertaintyLoad} \label{UncertaintyLoad}
\begin{align}
\mathcal{D}= \Bigg \{ d \in \mathbb{R}^{|N^{ac}|} : &  d_i \in [\bar{d}_i, \bar{d}_i+\gamma^D_i\hat{d}_i] \hspace{0.5cm}  \forall i \in N^{ac},  \\
& \sum_{i\in N^{ac}} \gamma^D_i \leq \Gamma^D, \\
& \gamma^D_i \in \{0,1\} \hspace{0.5cm}  \forall i \in N^{ac} \Bigg \}
\end{align}
\end{equationseries}

\begin{equationseries}{UncertaintyRES} \label{UncertaintyRES}
\begin{align}
\mathcal{P}= \Bigg \{ P \in \mathbb{R}^{|G^{res}|} : &  P_g \in [\bar{P}_g-\gamma^P_g\hat{P}_g, \bar{P}_g] \hspace{0.5cm}   \forall g \in G^{res},  \\
& \sum_{g\in G^{res}} \gamma^P_g \leq \Gamma^P,\\
& \gamma^P_g \in \{0,1\} \hspace{0.5cm}  \forall g \in  G^{res} \Bigg \}
\end{align}
\end{equationseries}


\begin{equationseries}{UncertaintyLine} \label{UncertaintyLine}
\begin{align}
\mathcal{F}= \Bigg \{ F \in  \mathbb{R}^{|L|} : &  F_l \in [\bar{F}_l-\gamma^F_l\hat{F}_l, \bar{F}_l] \hspace{0.5cm}  \forall l \in L, \\
& \gamma^F_l \geq  \gamma^{CG}_b \hspace{0.5cm}  \forall b \in  CG, l \in L_b,  \label{lineConv}\\
& \sum_{b \in CG} \gamma^{CG}_b \leq \Gamma^F,\\
& \gamma^{CG}_b \in \{0,1\} \hspace{0.5cm}  \forall b \in  CG, \\
& \gamma^{F}_l \in \{0,1\} \hspace{0.5cm}  \forall l \in  L \Bigg \},
\end{align}
\end{equationseries}

For the given first-stage decisions and uncertainty realisation, the recourse problem determines the minimum cost balancing actions with respect to energy balance, capacity, and frequency security constraints. Accordingly, the second-stage problem can be formulated as follows:

\begin{equationseries}{secondStage} \label{second} 
\begin{align}
\text{ min } & \sum_{g \in G^{con}} c^{act}_gr^{act}_g + \sum_{g \in G^{res}} (c^{up}_gp^{up}_g+c^{down}_gp^{down}_g) + \sum_{i \in N^{ac}} VOLLs^b_i  \label{second_obj}\\
\text{ s.t. } &  x^b_z-\left[ \sum_{i \in N_z^{ac}} \left(\sum_{g \in G^{con}_i} r^{act}_g + \sum_{g \in G^{res}_i} (p^{up}_g-p^{down}_g) - \sum_{(i,j) \in L} f^b_{ij}  +s^b_i \right) \right] \nonumber \\
& \hspace{7cm} =\sum_{i \in N_z^{ac}} \left(\sum_{g \in G_i} p_g  -  d_i +s_i \right)  \hspace{0.5cm} \forall  z \in Z \hspace{0.5cm} (\beta^{ac}_z) \label{second_ACzonal}\\
&  \sum_{i \in N_z^{dc}} \left(  \sum_{(j,i) \in L} f^{b}_{ji}- \sum_{(i,j) \in L} f^{b}_{ij} \right) = 0  \hspace{1cm} \forall  z \in Z  \hspace{0.5cm}  (\beta^{dc}_z) \label{second_DCzonal}\\
&  \sum_{z \in Z} x^b_z =0  \hspace{0.5cm}  (\theta) \label{second_marketClearing}\\
& x^b_z = 0 \hspace{1cm} \forall z \in Z^{owf} \hspace{0.5cm}  (\sigma_z) \label{second_obz_netzero} \\
&  f_l^{b,ac} - \sum_{z \in Z} ZPTDF_{lz}x^b_z = 0 \hspace{1cm} \forall  l \in CL  \hspace{0.5cm}  (\beta_l) \label{second_CL} \\
& -r^{act} \geq -r_g  \hspace{1cm} \forall  g \in G^{con} \hspace{0.5cm}  (m^{con}_g) \label{second_reserve}\\
& -p^{up}_g+p^{down}_g \geq -P_g + p_g \hspace{1cm} \forall  g \in G^{res} \hspace{0.5cm}  (m^{res}_g) \label{second_res}\\
& -s^b_i \geq -d_i +s_i \hspace{1cm} \forall  i \in N^{ac} \hspace{0.5cm}  (m_i) \label{second_demandShedding} \\
& f_l^{b,ac} \geq -F^{ac}_{l} \hspace{1cm} \forall l \in  CL \hspace{0.5cm}  (n^{ac}_l) \label{second_AClower}\\
& -f_l^{b,ac} \geq -F^{ac}_{l} \hspace{1cm} \forall l \in  CL \hspace{0.5cm}  (\nu^{ac}_l) \label{second_ACupper}\\
& f_l^{b} \geq -F_l \hspace{1cm} \forall l \in  L \hspace{0.5cm}  (n_l) \label{second_DClower}\\
& -f_l^{b} \geq -F_l \hspace{1cm} \forall l \in  L \hspace{0.5cm}  (\nu_l) \label{second_DCupper}\\
& -\delta^+_l - f_l^b \geq -F_l  \hspace{1cm} \forall  l \in L \hspace{0.5cm}  (\pi^+_l) \label{second_positiveHeadroom}\\
& -\delta^-_l + f_l^b \geq -F_l  \hspace{1cm} \forall  l \in L \hspace{0.5cm}  (\pi^-_l) \label{second_negativeHeadroom}\\
& \Delta T_{dc}^{ffr}\Bigg( \sum_{l \in L_s^{in}} \delta_l^+ + \sum_{l \in L_s^{out}} \delta_l^- \Bigg) \geq -2E_sK + \Bigg(\sum_{l \in L_s^{in}} \gamma_l^F f_l- \sum_{l \in L_s^{out} } \gamma_l^F f_l \Bigg)  \Delta T^{ffr} \hspace{0.5cm} \forall  s \in S \hspace{0.5cm}  (\mu_s^+) \label{second_underFrequency}\\
& \Delta T_{dc}^{ffr}\Bigg( \sum_{l \in L_s^{in}} \delta_l^- + \sum_{l \in L_s^{out}} \delta_l^+ \Bigg) \geq -2E_sK - \Bigg(\sum_{l \in L_s^{in}} \gamma_l^F f_l- \sum_{l \in L_s^{out} } \gamma_l^F f_l \Bigg)  \Delta T^{ffr} \hspace{0.5cm} \forall  s \in S \hspace{0.5cm}  (\mu_s^-) \label{second_overFrequency}\\
& x^b_z \text{ is free} \hspace{1cm} \forall z \in Z \\
& s^b_i \geq 0 \hspace{1cm} \forall i \in  N^{ac}  \label{second_demandShedDomain}\\
& r^{act}_g \geq 0 \hspace{1cm} \forall g \in G^{con} \\
& p^{up}_g, p^{down}_g \geq 0 \hspace{1cm} \forall g \in G^{res} \\
& \delta^+_l, \delta^-_l  \geq 0 \hspace{1cm} \forall l \in L, \label{second_domain}
\end{align}
\end{equationseries}
where the corresponding dual variables for the corresponding constraints, to be later used for the dual transformation, are given in the parentheses. 

The objective function \eqref{second_obj} minimises the cost of recourse actions taken to preserve the balance and keep frequency deviations within the limits. Constraints \eqref{second_ACzonal}-\eqref{second_obz_netzero} are similar to constraints \eqref{first_ACzonal}-\eqref{first_obz_netzero}, respectively. Constraints \eqref{second_reserve} ensure that reserve can be activated if procured in the first stage. Constraints \eqref{second_res} ensure that the net redispatch of renewable generators does not exceed their realised generation capacity, i.e., upward adjustment minus downward curtailment is bounded by the available output. Constraints \eqref{second_demandShedding} set a limit for demand shedding. Constraints \eqref{second_AClower}-\eqref{second_DCupper} enforce the flows on AC and DC lines to stay within the limits. Note that when a DC line fails, we have $F_l=0$, which forces $f_l^b=0$ through the constraints \eqref{second_DClower} and \eqref{second_DCupper}, representing the physical consequence of the converter outage on the associated line status. 

Constraints \eqref{second_positiveHeadroom}-\eqref{second_overFrequency} are the frequency security constraints representing frequency constraints for the fast frequency response (FFR). These constraints ensure that, at the post-balancing point, the power system can survive the instantaneous power loss from any tripped HVDC line connected to a converter. We use the energy-based linearisation, which bounds the maximum frequency deviation during the FFR phase. For each synchronous area $s \in S$, we define sets $L_s^{in}$ and $L_s^{out}$, representing the DC lines importing into and exporting from area $s$, respectively. The frequency parameters are as follows: nominal frequency $f_0=50$ Hz, maximum allowable deviation $\delta f^{max}= 0.8$ Hz, dead band $f_{db}=0$ Hz, FFR activation time $\Delta T^{ffr}=1.0$ s, HVDC ramp-up time $\Delta T^{dc}=0.5$ s. Based on these parameters, we derive the following constants $K=(\Delta f^{max}-f_{db})/f_0$ and HVDC FFR time factor $T^{ffr}_{dc}=\Delta T^{ffr} -\Delta T^{dc}/2$. 

Constraints \eqref{second_positiveHeadroom} and \eqref{second_negativeHeadroom} FFR headroom constraints where the emergency power available on each surviving DC line is limited by its remaining capacity after the balancing dispatch. Positive headroom represents the ability to increase flow on a line, injecting more power into the receiving area, whereas negative headroom represents the ability to decrease flow, reducing power leaving the sending area. Note that when a line $l$ is failed, $\delta^+_l=0$ and $\delta^-_l=0$, which means that the tripped line provides zero FFR. Additionally, the tripped line causes power deviation in both receiving and sending areas. Hence, the instantaneous power loss from tripped DC lines in area $s \in S$ is 
$$\Delta P_s^{trip}= \sum_{l \in L^{in}_s} \gamma_l^Ff_l-\sum_{l \in L_s^{out}} \gamma_l^Ff_l.$$
\noindent If $\gamma_l^F=1$, implying that line $l$ trips, then its flow $f_l$ is lost. The convention is that $\Delta P^{trip}_s > 0$ corresponds to a power deficit, because of lost imports, and $\Delta P^{trip}_s < 0$ corresponds to a surplus, because of lost exports. 

Constraints \eqref{second_underFrequency} and \eqref{second_overFrequency} are the frequency constraints for the FFR phase in case of power deficit and surplus, respectively, for each area $s$. If there is a power deficit, the frequency decreases. To limit the rate of change of frequency, specifically the rate of frequency decrease in area $s$, the FFR for a deficit comes from increasing imports and decreasing exports, implied by the following equation
$$\frac{\Delta P^{trip}_s \Delta T^{ffr} - \bigg(\sum\limits_{l \in L_s^{in}} \delta^+_l+ \sum\limits_{l \in L_s^{out}}  \delta_l^-\bigg) \Delta T^{ffr}_{dc}}{2E_s} \leq K.$$
\noindent If there is a power surplus, the frequency increases. To limit the rate of frequency increase in area $s$, the FFR for a surplus comes from decreasing imports and increasing exports, implied by the following equation
$$\frac{-\Delta P^{trip}_s \Delta T^{ffr} - \bigg(\sum\limits_{l \in L_s^{in}} \delta^-_l+ \sum\limits_{l \in L_s^{out}}  \delta_l^+\bigg) \Delta T^{ffr}_{dc}}{2E_s} \leq K.$$
Note that these frequency constraints do not harm the LP structure of the second-stage problem because they are linear for any fixed first-stage solution and uncertainty realisation. However, feasibility of the second-stage problem cannot be 
guaranteed for all combinations of first-stage decisions and uncertainty realisations, as the frequency constraints  \eqref{second_underFrequency}--\eqref{second_overFrequency} may become infeasible when insufficient inertia or FFR headroom 
is available. To ensure that the second-stage problem is always feasible, which is a prerequisite for applying the dual 
transformation, we introduce non-negative slack variables $\xi^+_s$ and $\xi^-_s$ for each synchronous area $s \in S$, 
relaxing the frequency constraints and updating them as follows:
\begin{align*}
& \Delta T_{dc}^{ffr}\Bigg( \sum_{l \in L_s^{in}} \delta_l^+ 
  + \sum_{l \in L_s^{out}} \delta_l^- \Bigg) + \xi^+_s 
  \geq -2E_sK + \Delta P^{trip}_s \Delta T^{ffr} 
  \hspace{0.5cm} \forall s \in S \\
& \Delta T_{dc}^{ffr}\Bigg( \sum_{l \in L_s^{in}} \delta_l^- 
  + \sum_{l \in L_s^{out}} \delta_l^+ \Bigg) + \xi^-_s 
  \geq -2E_sK - \Delta P^{trip}_s \Delta T^{ffr} 
  \hspace{0.5cm} \forall s \in S
\end{align*}
These slacks are penalised in the objective \eqref{second_obj} by a sufficiently large coefficient $M^{freq} \gg VOLL$, where the term $\sum\limits_{s \in S} M^{freq}(\xi^+_s+\xi^-_s)$ is added to the objective, ensuring that any solution activating the slacks is strictly dominated by solutions that satisfy 
the frequency constraints exactly. Hence, since the second-stage problem is LP, feasible, and bounded, the dual transformation method can be applied, which is essential for the proposed solution algorithm. The dual of problem \eqref{second} can be given as follows:

\begin{equationseries}{secondStageDual} \label{secondDual} 
\begin{align}
\text{ max } & \sum_{z \in Z}\beta^{ac}_z \sum_{i \in N_z^{ac}} \Big( \sum_{g \in G_i} p_g -d_i +s_i \Big) - \sum_{g \in G^{con}} m^{con}_g r_g - \sum_{g \in G^{res}} m^{res}_g (P_g-p_g)   \nonumber\\
& - \sum_{i \in N^{ac}} m_i (d_i-s_i) - \sum_{l \in CL} (n^{ac}_l+ \nu^{ac}_l) F^{ac}_l - \sum_{l \in L} (n_l + \nu_l)F_l  - \sum_{l \in L} (\pi^+_l + \pi^-_l)F_l  \nonumber \\
& - \sum_{s \in S} \mu^+_s \Big(2E_sK-\Delta P_s^{trip}\Big)- \sum_{s \in S} \mu^-_s \Big(2E_sK+\Delta P_s^{trip}\Big)\label{dual_obj}\\
\text{ s.t. } &  -\beta_{z_g}^{ac}-m^{con}_g \leq c^{act}_g \hspace{0.5cm} \forall  g \in G^{con} \hspace{0.5cm} (r^{act}_g) \label{dual_first}\\
&  -\beta_{z_g}^{ac}-m^{res}_g \leq c^{up}_g \hspace{0.5cm} \forall  g \in G^{res} \hspace{0.5cm} (p^{up}_g)\\
&  \beta_{z_g}^{ac}+m^{res}_g \leq c^{down}_g \hspace{0.5cm} \forall  g \in G^{res} \hspace{0.5cm} (p^{down}_g) \\
&  -\beta_{z_i^{ac}}^{ac}-m_i \leq VOLL \hspace{0.5cm} \forall  i \in N^{ac} \hspace{0.5cm} (s^{b}_i) \\
&  \beta_{z^{ac}_{l_1}}^{ac}+\beta_{z^{dc}_{l_2}}^{dc} +n_{l}-\nu_{l} - \pi^+_l + \pi^-_l = 0 \hspace{0.5cm} \forall  l \in L: l_1 \in N^{ac} \hspace{0.5cm} (f^{b}_l) \\
& -\beta_{z_{l_1}^{dc}}^{dc}+\beta_{z_{l_2}^{dc}}^{dc}+n_{l}-\nu_{l}  - \pi^+_l + \pi^-_l = 0 \hspace{0.5cm} \forall  l \in L: l_1 \in N^{dc} \hspace{0.5cm} (f^{b}_l) \\
&  \beta_l+n^{ac}_{ij}-\nu^{ac}_{ij} = 0 \hspace{0.5cm} \forall  l \in CL \hspace{0.5cm} (f^{b,ac}_l) \\
&  \beta_z^{ac}- \sum_{l \in CL} ZPTDF_{lz}\beta_l+ \theta = 0 \hspace{0.5cm} \forall  z \in Z \setminus Z^{owf}  \hspace{0.5cm} (x^{b}_z) \\
&  \beta_z^{ac}- \sum_{l \in CL} ZPTDF_{lz}\beta_l+ \theta +\sigma_z = 0 \hspace{0.5cm} \forall  z \in Z^{owf} \hspace{0.5cm} (x^{b}_z) \\
&  -\pi^+_l+\Delta T^{ffr}_{dc}(\mu^+_{s^{in}_l}+\mu^-_{s^{out}_l}) \leq 0 \hspace{0.5cm} \forall  l \in L \hspace{0.5cm} (\delta^+_l) \\
&  -\pi^-_l+\Delta T^{ffr}_{dc}(\mu^+_{s^{out}_l}+\mu^-_{s^{in}_l}) \leq 0 \hspace{0.5cm} \forall  l \in L \hspace{0.5cm} (\delta^-_l) \\
& \beta_z^{ac}, \beta_z^{dc} \text{ is free} \hspace{1cm} \forall  z \in Z\\
& \theta \text{ is free} \\
& \sigma_z \text{ is free} \hspace{1cm} \forall z \in Z^{owf} \\
& \beta_l \text{ is free} \hspace{1cm} \forall  l \in CL\\
& m_g^{con} \geq 0  \hspace{1cm} \forall g \in  G^{con} \\
& m_g^{res} \geq 0  \hspace{1cm} \forall g \in  G^{res} \\
& m_i \geq 0  \hspace{1cm} \forall i \in  N^{ac} \\
& n_l, \nu_l,  \pi^+_l,  \pi^-_l \geq 0 \hspace{1cm} \forall l \in  L \\
& n^{ac}_l, \nu^{ac}_l \geq 0 \hspace{1cm} \forall l \in  CL \\
& M^{freq} \geq  \mu_s^+, \mu_s^- \geq 0 \hspace{1cm} \forall s \in  S, \label{dual_domain}
\end{align}
\end{equationseries}
\noindent where $z_g$ is the zone in which generator $g \in G$ is located,  $z^{ac}_i$ is the zone in which AC node $i \in N^{ac}$ is located, $z^{dc}_i$ is the zone in which DC node $i \in N^{dc}$ is located, $s^{in}_l$ is the area to which line $l \in L$ comes, and $s^{out}_l$ is the area from which line $l \in L$ departs.

\subsection{Column-and-constraint generation algorithm}

We employ a Column-and-Constraint Generation (CCG) algorithm to solve an adaptive robust optimisation model (\cite{zeng2013solving}, \cite{sun2021robust}, \cite{damanik2025mitigating}). The main idea of CCG algorithm is to start with a relaxed master problem and update this master problem by discovered scenarios and the corresponding recourse variables and constraints iteratively until the convergence is reached. In this regard, the master problem and the oracle, which is the algorithm to detect the new scenarios, are constructed.

When the dual transformation is applied to the most inner problem, by strong duality, the three-level problem \eqref{three_level} can be rewritten as a bilevel problem such as 
\begin{equation}
\label{bilevel}
\min_{x^1 \in X^1} \; c_1(x^1)  \max_{u \in U, \lambda \in \Lambda(x^1,u)} b(x^1,u)\lambda,
\end{equation}
\noindent  where $\lambda$ represents the array of of dual variables and $b(x^1,u)$ represents the array of RHS' associated to constraints of the recourse problem. In this regard, the master problem and the oracle can be constructed.

\paragraph{Master problem:} At iteration $k$, the master problem is
\begin{equationseries}{master} \label{master}
\begin{align}
\min_{\begin{array}{c}
\alpha,f^{ac}, f, x, E, s,r,p, \eta\\
\{{f^{b,ac}}^j, {f^{b}}^j,{x^{b}}^j,{s^{b}}^j, {r^{act}}^j, {p^{{up}}}^j, \\ \hspace{1cm}  {p^{down}}^j,{\delta^{+}}^j, {\delta^{-}}^j, {\xi^+}^j, {\xi^-}^j\}_{j=1}^{k}
\end{array}} \quad & \sum_{g \in G} c_gp_g + \sum_{g \in G^{con}} (C_g\alpha_g+ c^r_gr_g)  + \sum_{i \in N^{ac}} VOLLs_i  + \eta  \\
\text{s.t.} \quad & (\alpha,f^{ac}, f, x, E, s,r,p) \in X^1 \\
& \eta \geq \sum_{g \in G^{con}} c^{act}_g{r^{act}}^j_g + \sum_{g \in G^{res}} (c^{up}_g{p^{up}}^j_g+c^{down}_g{p^{down}_g}^j) \nonumber \\
& \hspace{0.5cm} +  \sum_{i \in N^{ac}} VOLL{s^b_i}^j + \sum_{s \in S} M^{freq}({\xi^+_s}^j + {\xi^-_s}^j) \hspace{0.5cm} \forall\, j = 1, \ldots, k \\
& ({f^{b,ac}}^j, {f^{b}}^j,{x^{b}}^j,{s^{b}}^j, {r^{act}}^j, {p^{{up}}}^j, {p^{down}}^j,{\delta^{+}}^j, {\delta^{-}}^j) \nonumber \\
& \hspace{3.5cm} \in X^2(d^j, P^j, F^j) \hspace{0.5cm} \forall\, j = 1, \ldots, k \\
& \eta \geq 0, \quad  \label{eq:master_cuts}
\end{align}
\end{equationseries}
where $X^1$ represents the set of constraints \eqref{first_ACzonal}-\eqref{first_domainGeneration},  $X^2(d^j, P^j, F^j)$ represents the set of recourse constraints \eqref{second_ACzonal}-\eqref{second_domain} under the scenario $(d^j, P^j, F^j) \in \mathcal{D} \times \mathcal{P} \times \mathcal{F}$, and $\eta$ denotes the maximum of recourse costs, which are evaluated for a subset of scenarios. Let $x^k=(\alpha^k,{f^{ac}}^k, f^k, x^k, E^k, s^k,r^k,p^k)$ be the optimal solution and $z^k$ be the optimal objective function value of problem \eqref{master}.

\paragraph{Oracle:} Given the master solution $x^k$, the following problem is solved to identify the worst-case scenario.
\begin{equationseries}{oracle} \label{oracle}
\begin{align}
\max_{\begin{array}{c}
d,P,F, \gamma^D, \gamma^P, \gamma^F, \gamma^{CG}\\
\beta^{ac},\beta^{dc}, \theta, \sigma, \beta, m^{con}, m^{res}, m\\
n^{ac}, \nu^{ac}, n, \nu,\pi^{+}, \pi^{-},\mu^{+}, \mu^{-} \end{array}} \quad & \sum_{z \in Z}\beta^{ac}_z \sum_{i \in N_z^{ac}} \Big( \sum_{g \in G_i} {p_g}^k -d_i + {s_i}^k \Big) - \sum_{g \in G^{con}} m^{con}_g {r_g}^k - \sum_{g \in G^{res}} m^{res}_g (P_g-{p_g}^k)   \nonumber\\
& - \sum_{i \in N^{ac}} m_i (d_i-{s_i}^k) - \sum_{l \in CL} ({n^{ac}_l}+ {\nu^{ac}_l}) F_l^{ac} - \sum_{l \in L} ({n_l} + {\nu_l} +{\pi^+_l} + {\pi^-_l})F_l   \nonumber \\
& - \sum_{s \in S} \left[(\mu^+_s + \mu^-_s) 2{E_s}^kK - (\mu^+_s - \mu^-_s) \left(\sum_{l \in L^{in}_s} \gamma_l^F{f_l}^k-\sum_{l \in L_s^{out}} \gamma_l^F{f_l}^k \right)\right] \label{oracle_obj}\\
\text{s.t.} \quad & (\beta^{ac}, \beta^{dc}, \theta,\beta, m^{con}, m^{res}, m,
n^{ac}, \nu^{ac}, n, \nu,\pi^{+}, \pi^{-},\mu^{+}, \mu^{-}) \in \Lambda \\
& (d, \gamma^D) \in \mathcal{D} \\
& (P, \gamma^P) \in \mathcal{P} \\
& (F, \gamma^F, \gamma^{CG}) \in \mathcal{F},  \label{eq:oracle_feasSets}
\end{align}
\end{equationseries}
where $\Lambda$ represents the set of constraints \eqref{dual_first}-\eqref{dual_domain}. 
The objective function 
\eqref{oracle_obj} contains bilinear terms arising from the product 
of dual variables and uncertainty set variables. Since only the 
objective function is nonlinear, the optimal solution still occurs 
at the extreme points of the feasible region, so the validity of 
the generated cuts is not affected (\cite{sun2021robust}). 
Nevertheless, solving a nonlinear problem repeatedly can be 
time-consuming. To obtain a mixed-integer linear program, we 
substitute the uncertain parameters by their definitions, i.e., 
$d_i = \bar{d}_i + \hat{d}_i\gamma^D_i$, 
$P_g = \bar{P}_g - \hat{P}_g\gamma^P_g$, and 
$F_l = \bar{F}_l - \hat{F}_l\gamma^F_l$, which produces bilinear terms of the form continuous dual variable 
times binary uncertainty indicator, such as $\beta^{ac}_z \cdot \gamma^D_i$ 
and $m_i \cdot \gamma^D_i$ from demand uncertainty, 
$m^{res}_g \cdot \gamma^P_g$ from RES uncertainty, and 
$n_l \cdot \gamma^F_l$, $\nu_l \cdot \gamma^F_l$, 
$\pi^+_l \cdot \gamma^F_l$, $\pi^-_l \cdot \gamma^F_l$, 
$\mu^+_s \cdot \gamma^F_l$, and $\mu^-_s \cdot \gamma^F_l$ 
from line failure uncertainty.
We linearise these products exactly using McCormick 
envelopes (\cite{mccormick1976computability}). For each product 
$w = \lambda \cdot \gamma$ where 
$\lambda \in [\lambda^L, \lambda^U]$ and $\gamma \in \{0,1\}$, 
the following four linear inequalities provide an exact 
representation:
\begin{align}
w &\leq \lambda^U \gamma, \nonumber\\
w &\geq \lambda^L \gamma, \nonumber\\
w &\leq \lambda - \lambda^L(1-\gamma), \nonumber\\
w &\geq \lambda - \lambda^U(1-\gamma). \nonumber
\end{align}
Since one of the two factors is binary, the McCormick relaxation is 
exact, i.e., it introduces no approximation error, and the oracle 
reduces to a mixed-integer linear program. The bounds $\lambda^L$ 
and $\lambda^U$ on the dual variables are set to a sufficiently 
large value $M$, which is selected by verifying that it is not 
binding at optimality for any of the instances tested in the 
computational experiments.

Let $(d^{k+1}, P^{k+1}, F^{k+1})$ be the optimal solution and $o^{k+1}$ be the optimal objective function value of problem \eqref{oracle}.
Accordingly, the complete solution procedure for the two-stage robust problem is given in Algorithm~\ref{alg:ccg}.
 
\begin{algorithm}[htbp]
\caption{Column-and-Constraint Generation}
\label{alg:ccg}
\begin{algorithmic}[1]
\STATE Set $LB \leftarrow -\infty$, $UB \leftarrow +\infty$, $k \leftarrow 0$
\REPEAT
    \STATE Solve master problem \eqref{master}: Obtain $x^k$
    \STATE Solve oracle problem \eqref{oracle} for given $x^k$: Obtain scenario $(d^{k+1}, P^{k+1}, F^{k+1})$
    \STATE $LB \leftarrow z^k$
    \STATE $UB \leftarrow \min\left(UB, \sum\limits_{g \in G} c_gp^k_g + \sum\limits_{g \in G^{con}} (C_g\alpha^k_g+ c^r_gr^k_g)  + \sum\limits_{i \in N^{ac}} VOLLs^k_i  + o^{k+1}\right)$
    \IF{$(UB - LB)/UB \leq \epsilon$}
        \STATE \textbf{stop}
    \ENDIF
    \STATE Add columns $({f^{b,ac}}^{k+1}, {f^{b}}^{k+1},{x^{b}}^{k+1},{s^{b}}^{k+1}, {r^{act}}^{k+1}, {p^{{up}}}^{k+1}, {p^{down}}^{k+1},{\delta^{+}}^{k+1}, {\delta^{-}}^{k+1})$ and corresponding constraints based on scenario $(d^{k+1}, P^{k+1}, F^{k+1})$ to the master problem
    \STATE $k \leftarrow k + 1$
\UNTIL{maximum iterations reached}
\end{algorithmic}
\end{algorithm}

\section{Computational experiments} \label{sec:case}
In this section, we explain the details of the data related to the system considered  and present the computational study comparing the home market (HM) and offshore bidding zone (OBZ) designs under different levels of uncertainty.

\subsection{Details of the system data}

\begin{figure}[h!]
	\centering
	\includegraphics[trim=150 550 1000 0,width=\textwidth]{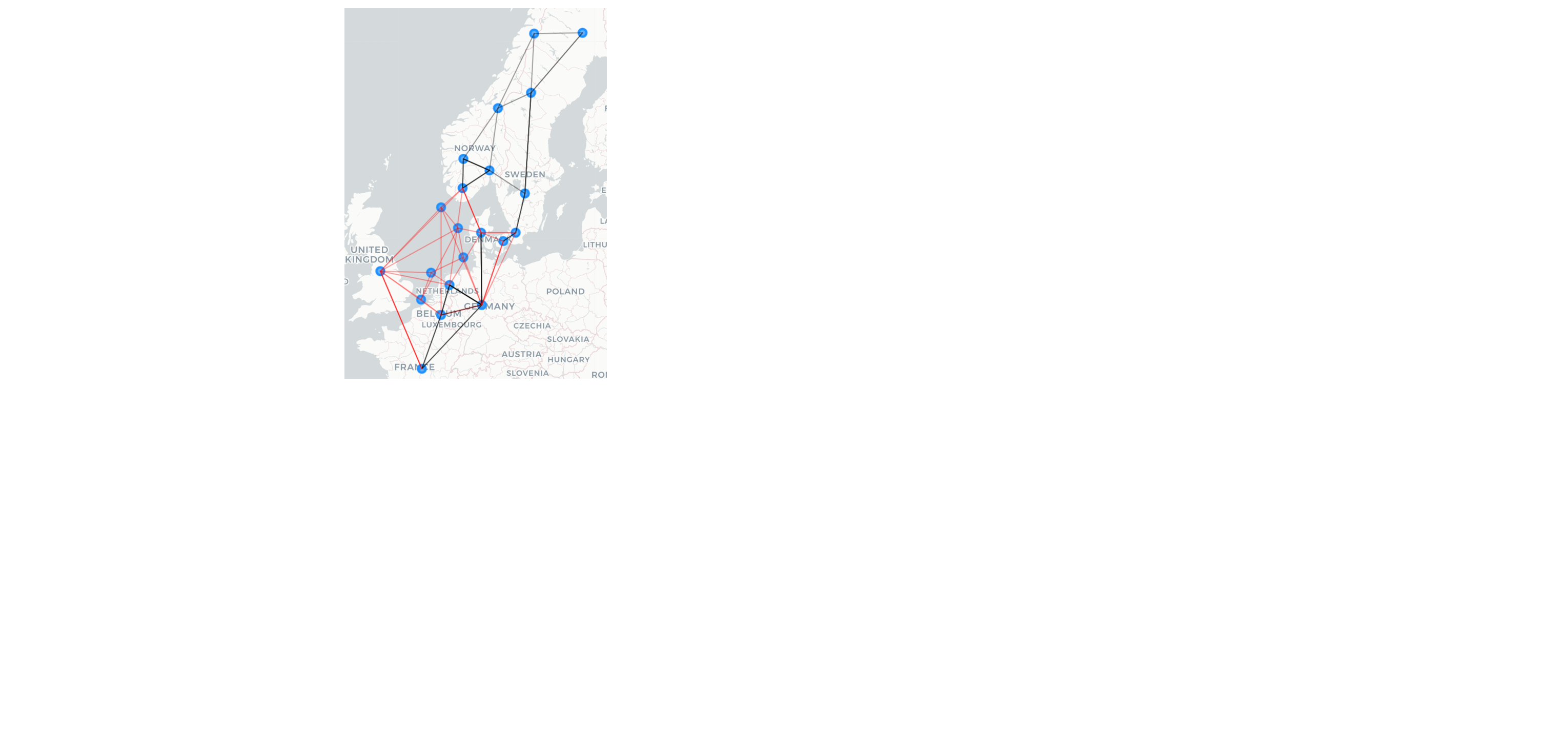}
	\caption{North Sea grid}
	\label{fig:north_sea}
\end{figure}

The proposed model is applied to a North Sea AC/DC test system. The system represents the high-voltage transmission grids of 16 bidding zones across three synchronous areas, which are Continental Europe, the Nordic system, and Great Britain, interconnected by an extensive HVDC network including five offshore wind farm hubs connected through a meshed offshore DC grid. Table~\ref{tab:system_summary} summarises the key characteristics of the North Sea grid depicted in Figure \ref{fig:north_sea}, where bidding zones and OWF hubs are represented in blue dots and DC and AC connections are represented in red and black, respectively.

\begin{table}[htbp]
\centering
\caption{Characteristics of the North Sea grid}
\label{tab:system_summary}
\small
\begin{tabular}{@{}lr@{}}
\toprule
Component & Count \\
\midrule
AC buses \hspace{3cm}  & 3{,}406 \\
AC lines & 4{,}834 \\
Inter-zonal AC lines & 59 \\
DC buses  & 48 \\
DC lines & 42 \\
AC/DC converters & 48 \\
Generators  & 3{,}945 \\
Loads & 1{,}767 \\
Offshore wind farm hubs & 5 \\
Synchronous areas  & 3 \\
\bottomrule
\end{tabular}
\end{table}

The three synchronous areas are defined as follows. The Continental European area comprises Belgium (BE), Germany (DE), Western Denmark (DK1), France (FR), and the Netherlands (NL). The Nordic area comprises Eastern Denmark (DK2), Norway (NO1--NO5), and Sweden (SE1--SE4). Note that Western Denmark (DK1) is AC-connected to Germany and therefore belongs to the Continental synchronous area, while Eastern Denmark (DK2) is AC-connected to Sweden and belongs to the Nordic area. Great Britain (UK) forms its own synchronous area, connected to the continent and Nordic system exclusively through HVDC links.

The grid includes five offshore wind farm hubs with a total installed capacity of 29.5~GW, as listed in Table~\ref{tab:owf}. Each hub is connected to the onshore grid through multiple HVDC lines and to other hubs through a meshed offshore DC grid, proposed by reports of \cite{OTC_2} and \cite{OTC_4}. Power flow on the converter stations connecting offshore  wind farms to their DC hubs is considered unidirectional, as these platforms have no load or AC connections that could absorb imported power. The AC network and generation fleet, including locations, capacities, and technology types, are sourced from the Ten Year Network Development Plan 2024 National Trends scenario \cite{TYNDP2024} and \cite{EUGO}. We consider the technology types of solar PV, onshore wind, and offshore wind in the set of $G^{res}$, where their maximum generation capacities are calculated based on their installed capacities and capacity factors, which change depending on zone, season, and hour of a day. The capacity factors are derived per zone based on actual generation and installed capacity in 2025 per production type data in \cite{REScapFac}. If there is not available data for a zone, the capacity factor is adjusted based on the assumption that it decreases for solar and increases for wind toward the north. Additionally, for hydro pumped storage, which constitutes an important part of Norway's electricity supply, we consider a seasonal capacity factor where the capacity factor is higher in summer due to high inflow (\cite{hydroPump}). While inertia and marginal costs of conventional generators are based on \cite{EUGO}, marginal costs of renewable generators are set to a small cost of $1$ EUR/MWh. For no-load costs of conventional generators, a coefficient is calculated based on \cite{schroder2013current} per unit of capacity of each technology type.

\begin{table}[htbp]
\centering
\caption{Offshore wind farm hubs}
\label{tab:owf}
\small
\begin{tabular}{@{}lrll@{}}
\toprule
Hub  & Capacity (GW) & Zone (under HM) & Zone (under OBZ) \\
\midrule
Princess Elisabeth Island & 3.5 & $BE$ & $OBZ_{BE}$ \\
NL Energy Hubs & 10& $NL$ & $OBZ_{NL}$  \\
German Offshore Cluster & 10 & $DE$ & $OBZ_{DE}$ \\
Danish Energy Island & 3 & $DK1$ & $OBZ_{DK}$  \\
S\o{}rlige Nordsj\o{} 2 & 3 & $NO2$ & $OBZ_{NO}$  \\
\bottomrule
\end{tabular}
\end{table}

The AC bus information with the associated loads and  characteristics of AC lines is sourced from \cite{EUGO}. 
$NPTDF$ and $ZPTDF$ matrices are derived according to characteristics of AC lines. Note that, although $NPTDF$ is the same for both designs, $ZPTDF$ differs because of the change in zone configurations. Line susceptance ($B$) of an AC line with endpoint buses $i$ and $j$, is calculated based on resistance ($R$) and reactance ($X$) (\cite{den2014dc}).


$$B_{ij}= \frac{X_{ij}}{R_{ij}^2+X_{ij}^2}$$

\noindent The $NPTDF$ matrix can then be
calculated as follows (\cite{den2014dc}):

$$NPTDF^{|L^{ac}| \times |N^{ac}|}= (B_dA)(A^TB_dA)^{-1},$$
where $B_d$ is the diagonal matrix ($|L^{ac}| \times |L^{ac}|$) of line susceptances and $A$ is the incidence matrix ($|L^{ac}| \times |N^{ac}|$ with +1 for the sending bus and -1 for the receiving bus). This follows from the DC power flow approximation
$$P_{ij}=B_{ij}(\theta_i-\theta_j),$$
$$P_i=\sum_{j \in N: (i,j) \in E^{ac}} B_{ij}(\theta_i-\theta_j) + \sum_{j \in N: (j,i) \in E^{ac}} B_{ji}(\theta_j-\theta_i),$$
after setting the angles of reference nodes to 0. 






Finally, after NPTDF is reduced to the rows corresponding to critical inter-zonal lines, $ZPTDF$ is calculated as follows:

$$ZPTDF^{|CL| \times |Z|}= NPTDF^{|CL| \times |N^{ac}|} \cdot GSK^{|N^{ac}| \times |Z|},$$
where $GSK$ is the generation shift key matrix, which maps one unit increase in zonal net injection to the amount of increase or decrease in nodal injections across all buses in that zone. To calculate $GSK$, the flat strategy is used (\cite{risanger2025flexible}). 

\subsection{Design of experiments}
We design computational experiments to address the comparison between HM and OBZ designs under robust market clearing and investigate the effects of uncertainty levels on the results obtained from both designs. For the different uncertainty levels, different settings of $\Gamma$ values are considered and presented in Table \ref{tab:gamma_levels}. 
The converter contingency budget $\Gamma^F = 1$ implies that one converter station may trip, thereby isolating its associated DC bus and all connected lines. This reflects a realistic N-1 security criterion for HVDC systems. 
For the RES uncertainty, a pure uncertainty setting is employed. In the North Sea network, simultaneous deviations from forecasted generation across multiple renewable sources are plausible, particularly under widespread weather events. Therefore, the generation uncertainty budget $\Gamma^P$ is set to its maximum value. 
However, simultaneous deviations at all load centers are considerably less likely. Modeling load uncertainty under a pure uncertainty assumption would therefore be overly conservative and would lead to unrealistically high system costs. For this reason, the demand uncertainty budget $\Gamma^D$ is varied to represent a range of conditions, from perfect forecasts to widespread forecast errors.

\begin{table}[htbp]
\centering
\caption{Uncertainty budget settings}
\label{tab:gamma_levels}
\small
\begin{tabular}{@{}lccccc@{}}
\toprule
Level & $\Gamma^D$ & $\Gamma^P$ & $\Gamma^F$ & Description \\
\midrule
$\Gamma_0$ & 0 & 0 & 0 & Deterministic \\
$\Gamma_1$ & 0 & 0 & 1 & Converter contingency only \\
$\Gamma_2$ & 0 & 3060 & 1 & Contingency and pure uncertainty for RES  \\
$\Gamma_3$ &  89 & 3060 & 1 & Contingency, RES, and uncertainty for 5\% of load\\
$\Gamma_4$ & 177 & 3060 & 1 & Contingency, RES, and uncertainty for 10\% of load\\
$\Gamma_5$ & 266 & 3060 & 1 & Contingency, RES, and uncertainty for 15\% of load\\
\bottomrule
\end{tabular}
\end{table}

\begin{figure}[htbp]
\centering
\begin{minipage}{0.49\textwidth}
\centering
{Home market (HM)}\\
{\footnotesize OWF joins zone A $\rightarrow$ 2 zones}\\[2mm]
\begin{tikzpicture}[
    every node/.style={font=\footnotesize},
    bus/.style={draw, rounded corners=2pt, minimum width=14mm, minimum height=7mm, align=center},
    busA/.style={bus, fill=blue!10, draw=blue!60},
    busB/.style={bus, fill=red!10, draw=red!60},
    owf/.style ={bus, fill=green!12, draw=green!55!black},
    dc/.style  ={bus, fill=teal!10, draw=teal!70!black},
    zoneA/.style={draw=blue!60, dashed, rounded corners=4pt, inner sep=2.5mm},
    zoneB/.style={draw=red!60, dashed, rounded corners=4pt, inner sep=2.5mm},
    aclink/.style={-{Stealth}, thick, black!55!black},
    dclink/.style={-{Stealth}, thick, red!55!black},
    convlink/.style={-{Stealth}, thick, green!55!black, dashed},
]
\node[busA] (b1)   at (0,    0)    {Bus 1};
\node[busA] (b2)   at (0,   -1)    {Bus 2};
\node[dc]   (dc10) at (2.4, -1)    {DC 10};
\node[owf]  (b4)   at (0,   -2.7)  {Bus 4};
\node[below=-1pt of b4, font=\scriptsize] {OWF gen};
\node[dc]   (dc40) at (2.4, -2.7)  {DC 40};
\node[busB] (b3)   at (0,   -4.6)  {Bus 3};
\node[dc]   (dc30) at (2.4, -4.6)  {DC 30};

\begin{scope}[on background layer]
\node[zoneA, fit=(b1)(b2)(b4)(dc10)(dc40),
      label={[blue!60, font=\scriptsize]above:Zone A}] {};
\node[zoneB, fit=(b3)(dc30),
      label={[red!60, font=\scriptsize]left:Zone B}] {};
\end{scope}

\draw[dclink] (dc40) -- node[right, font=\scriptsize]{intra-zonal} (dc10);
\draw[aclink] (b1) -- node[right, font=\scriptsize]{AC} (b2);
\draw[convlink] (b4) -- node[above]{\footnotesize conv} (dc40);
\draw[convlink] (b1) -- node[above]{\footnotesize conv} (dc10);
\draw[convlink] (b3) -- node[above]{\footnotesize conv} (dc30);
\draw[dclink] (dc40.east) .. controls +(0.5,0) and +(0.5,0.5) .. (dc30.east)
    node[right, pos=0.55, font=\scriptsize]{inter-zonal};
\end{tikzpicture}
\end{minipage}
\hfill
\begin{minipage}{0.49\textwidth}
\centering
{Offshore bidding zone (OBZ)}\\
{\footnotesize OWF forms own zone $\rightarrow$ 3 zones}\\[2mm]
\begin{tikzpicture}[
    every node/.style={font=\footnotesize},
    bus/.style={draw, rounded corners=2pt, minimum width=14mm, minimum height=7mm, align=center},
    busA/.style={bus, fill=blue!10, draw=blue!60},
    busB/.style={bus, fill=red!10, draw=red!60},
    owf/.style ={bus, fill=green!12, draw=green!55!black},
    dc/.style  ={bus, fill=teal!10, draw=teal!70!black},
    zoneA/.style  ={draw=blue!60,        dashed, rounded corners=4pt, inner sep=2.5mm},
    zoneB/.style  ={draw=red!60,         dashed, rounded corners=4pt, inner sep=2.5mm},
    zoneOBZ/.style={draw=green!55!black, dashed, rounded corners=4pt, inner sep=2.5mm},
    aclink/.style={-{Stealth}, thick, black!55!black},
    dclink/.style={-{Stealth}, thick, red!55!black},
    convlink/.style={-{Stealth}, thick, green!55!black, dashed},
]
\node[busA] (b1)   at (0,    0)    {Bus 1};
\node[busA] (b2)   at (0,   -1)    {Bus 2};
\node[dc]   (dc10) at (2.4, -1)    {DC 10};
\node[owf]  (b4)   at (0,   -2.7)  {Bus 4};
\node[below=-1pt of b4, font=\scriptsize] {OWF gen};
\node[dc]   (dc40) at (2.4, -2.7)  {DC 40};
\node[busB] (b3)   at (0,   -4.6)  {Bus 3};
\node[dc]   (dc30) at (2.4, -4.6)  {DC 30};

\begin{scope}[on background layer]
\node[zoneA, fit=(b1)(b2)(dc10),
      label={[blue!60, font=\scriptsize]above:Zone A}] {};
\node[zoneOBZ, fit=(b4)(dc40),
      label={[green!55!black, font=\scriptsize]left:Zone OBZ}] {};
\node[zoneB, fit=(b3)(dc30),
      label={[red!60, font=\scriptsize]left:Zone B}] {};
\end{scope}

\draw[dclink] (dc40) -- node[right, font=\scriptsize]{inter-zonal} (dc10);
\draw[aclink] (b1) -- node[right, font=\scriptsize]{AC} (b2);
\draw[convlink] (b4) -- node[above]{\footnotesize conv} (dc40);
\draw[convlink] (b1) -- node[above]{\footnotesize conv} (dc10);
\draw[convlink] (b3) -- node[above]{\footnotesize conv} (dc30);
\draw[dclink] (dc40.east) .. controls +(0.35,0) and +(0.35,0.5) .. (dc30.east)
    node[right, pos=0.55, font=\scriptsize]{inter-zonal};
\end{tikzpicture}
\end{minipage}

\caption{Zone assignment under HM (left) and OBZ (right)}
\label{fig:zone_assignment}
\end{figure}
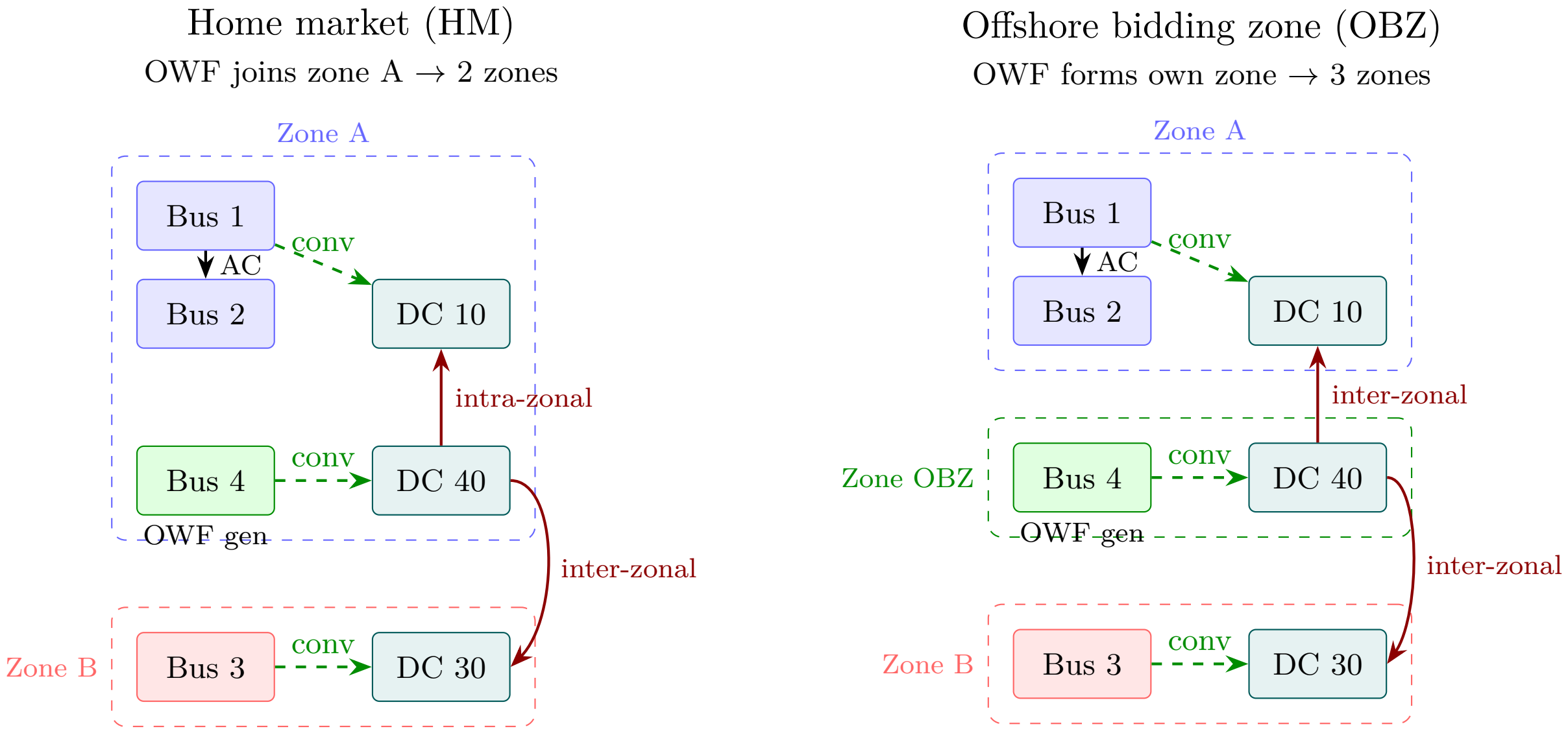

For the better understanding of structural differences between HM and OBZ, we give a small example network in Figure \ref{fig:zone_assignment}. Note that the physical infrastructure is identical in both designs, but the difference lies in the zone boundaries. Under HM, the offshore wind farm at bus 4 and its DC bus 40 are assigned to the home country zone A. The main HVDC line from DC 40 to DC 10 is intra-zonal and thus the day-ahead market treats it as internal to zone $A$ and does not model its capacity limit. Only the line from DC 40 to DC 30 crosses a zone boundary and is visible to the market as an inter-zonal flow. Consequently, if the converter at bus 4 trips and isolates DC 40, the market has no mechanism to anticipate the resulting loss of the intra-zonal line (DC 40$\to$DC 10), and the day-ahead dispatch may become physically undeliverable, requiring redispatch from generators at buses 1 or 2 or load shedding. Under OBZ, the offshore wind farm and its DC hub form a separate zone. Now both lines leaving DC~40 are inter-zonal. The market explicitly models both line capacities and can observe the congestion between the offshore hub and the onshore zones. If the converter at bus~4 trips, the market sees which inter-zonal flows are lost and can procure reserves accordingly. In short, the difference in zone boundaries has three consequences in the day-ahead market clearing. First, congestion on the main line DC~40$\to$DC~10 is invisible to the market under HM because it is intra-zonal while this line is inter-zonal and its capacity constraint appears explicitly in the zonal power balance equations under OBZ. Second, the offshore wind farm receives zone~$A$'s price regardless of line congestion under HM, but the offshore zone develops its own price that reflects the marginal cost of evacuating power through congested lines under OBZ. When evacuation capacity is scarce, the OBZ price drops below onshore prices, signaling the need for transmission investment. Third, under HM, a converter contingency at bus~4 affects only intra-zonal flows, so the frequency constraints cannot target reserves to cover this specific loss. Under OBZ, the same contingency disrupts inter-zonal flows that are explicitly modeled, enabling the market to procure targeted reserves and schedule HVDC fast frequency response on the surviving lines.

A common finding in the bidding zone literature is that the choice of zone configuration is welfare-neutral in deterministic settings, where a more granular zone design reduces redispatch but increases day-ahead cost by exactly the same amount, leaving the total (DA + redispatch) cost unchanged \cite{kenis2023off}. This result holds when the day-ahead and redispatch stages solve the same underlying economic dispatch with identical feasibility constraints. However, in the robust problem studied in this paper, having binary commitment decisions and frequency security constraints break this equivalence. First, the commitment decisions in the day-ahead stage determine which generators are available for redispatch, and different zone configurations lead to different commitment patterns, creating different feasibility regions for the subsequent redispatch actions. Second, the inertia and FFR constraints couple the commitment decisions with DC flow scheduling. The zonal market's visibility or lack of offshore lines directly affects how much inertia is procured and where FFR headroom is positioned. For these reasons, comparing the HM and OBZ designs solely on the basis of the total cost obtained from zonal planning would be misleading, since both the day-ahead cost and the cost of recourse actions may be higher under OBZ. Hence, using the optimal day-ahead decisions and the worst-case realisation of the uncertain parameters obtained from the proposed adaptive robust optimisation model, we perform a post-analysis stage. In this stage, the nodal redispatch problem \eqref{nodal_second} is solved, in which nodal injections \(x_i^{\text{nodal}}\), \(i \in N^{ac}\), and flows on each AC line \(f_l^{b,ac}\), \(l \in L^{ac}\), are modelled explicitly, so that any infeasibilities arising from the lack of physical considerations in zonal planning can be resolved. This evaluation approach ensures that a market design is not rewarded for ignoring physical constraints. If OBZ's additional zone boundaries force the day-ahead market to internalise offshore line congestion, raising DA costs, but this produces a dispatch that requires far less nodal correction, the true total cost including DA cost and nodal redispatch will correctly reflect the net benefit. Conversely, if HM achieves a lower DA cost by ignoring offshore congestion, the resulting nodal infeasibilities will appear as high redispatch cost, revealing the hidden cost of the market's information gap. Hence, in Section \ref{results}, we present the nodal dispatch costs as well.


\begin{equationseries}{NodalsecondStage} \label{nodal_second} 
\begin{align}
\text{ min } & \sum_{g \in G^{con}} c^{act}_gr^{act}_g + \sum_{g \in G^{idle}} (C_g u_g + \kappa^{emer} c_g p^{emer}_g) + \sum_{g \in G^{res}} (c^{up}_gp^{up}_g+c^{down}_gp^{down}_g) + \sum_{i \in N^{ac}} VOLLs^b_i  \label{nodal_second_obj}\\
\text{ s.t. } &  x^{nodal}_i- \left(\sum_{g \in G^{con}_i} r^{act}_g + \sum_{g \in G^{idle}_i} p^{emer}_g + \sum_{g \in G^{res}_i} (p^{up}_g-p^{down}_g) - \sum_{(i,j) \in L} f^b_{ij}  +s^b_i \right) \nonumber \\
&  \hspace{9cm} =\sum_{g \in G_i} p_g  -  d_i +s_i  \hspace{0.5cm} \forall  i \in N^{ac}  \label{nodal_second_ACzonal}\\
& p^{emer}_g \leq P^{max}_g u_g \hspace{1cm} \forall g \in G^{idle} \label{nodal_emer_cap}\\
&   \sum_{(j,i) \in L} f^{b}_{ji}- \sum_{(i,j) \in L} f^{b}_{ij} = 0  \hspace{1cm} \forall  i \in N^{dc} \label{nodal_second_DCzonal}\\
&  \sum_{i \in N^{ac}} x^{nodal}_i =0 \label{nodal_second_marketClearing}\\
&  f_l^{b,ac} - \sum_{i \in N^{ac}} NPTDF_{li}x^{nodal}_i = 0 \hspace{1cm} \forall  l \in L^{ac}   \label{nodal_second_Lac} \\
& f_l^{b,ac} \geq -F_l^{ac} \hspace{1cm} \forall l \in  L^{ac} \label{nodal_second_AClower}\\
& -f_l^{b,ac} \geq -F_l^{ac} \hspace{1cm} \forall l \in  L^{ac} \label{nodal_second_ACupper}\\
& x^{nodal}_i \text{ is free} \hspace{1cm} \forall i \in N^{ac} \\
& p^{emer}_g \geq 0 \hspace{1cm} \forall g \in G^{idle} \label{nodal_emer_domain_p}\\
& u_g \in \{0,1\} \hspace{1cm} \forall g \in G^{idle} \label{nodal_emer_domain_u}\\
& \eqref{second_reserve}-\eqref{second_demandShedding},\eqref{second_DClower}-\eqref{second_overFrequency}, \eqref{second_demandShedDomain} - \eqref{second_domain} 
\end{align}
\end{equationseries}

In the nodal redispatch problem, unlike the zonal redispatch,  we introduce an emergency generation option from the set of  generators $G^{idle}$. When physical network limitations are  fully enforced, load shedding can become unrealistically high  without this option. Since gas-fired generators typically have  short start-up times, they can be brought online at short notice. Hence, we define the set of generators eligible for emergency  dispatch as idle gas generators that were not committed in the day-ahead stage, i.e., $G^{idle} = \{g \in  G^{con}: \text{generator type is gas and } \alpha_g=0\}$. The emergency generation quantity is represented by $p^{emer}_g$ and the binary start-up indicator by $u_g$ for each $g \in G^{idle}$. Since this dispatch is unplanned, emergency generation must be more expensive than day-ahead generation. Accordingly, we apply an emergency cost multiplier $\kappa^{emer}$ to the marginal generation cost. In the computational experiments, we set 
$\kappa^{emer}$ to $3$. 

The computational experiments are performed on a 64-bit Windows machine equipped with an Intel(R) Core(TM) Ultra 7 165H processor and 64 GB RAM. The algorithm is coded in Julia 1.12.3 using Gurobi 13.0.0.  In the experiments, the cost of unit load shedding $VOLL$ is set to $1{,}000$ EUR/MWh. $M^{freq}$ is set to $10 \times VOLL$, where, in all instances tested, the slack variables are zero at optimality, confirming that the frequency constraints are satisfied without relaxation. The big $M$ value for McCormick envelopes is set to $10^6$. For the oracle, a time limit of 10 minutes is imposed. If an optimal solution cannot be found within the time limit, the best integer solution found until the termination is provided to the master problem. Additionally, the maximum number of iterations for the CCG algorithm is set to 20, and the convergence tolerance $\epsilon$ is set to $10^{-4}$. It can be said that all runs provided optimal robust solutions within 180 minutes, and the algorithm usually converged in fewer than 10 iterations. The source code and data used for the computational experiments can be accessed via the link \href{https://github.com/OzlemMahmutogullari/NorthSeaData}{https://github.com/OzlemMahmutogullari/NorthSeaData}. 

\subsection{Results and discussion} \label{results}

\subsubsection{Value of considering uncertainty}
To quantify the value of robustness, under different uncertainty levels, we compare performances of the deterministic and robust solutions. In this regard, we first solve the problem under no uncertainty $\Gamma$0, then employ the deterministic day-ahead actions in the worst-case scenarios identified by uncertainty levels $\Gamma$1-$\Gamma$5. Hence, we can investigate how valuable employing a robust solution can be in an uncertain environment because the deterministic solution is not able to cope with the risks in an efficient way. Table \ref{tab:operational} presents the reserve procurement, committed gas capacity, real-time load shedding in nodal redispatch, and the total inertia ensured, respectively, when deterministic and robust solutions are employed. Note that committed gas capacity is reported because gas generators are the technology whose committed capacity varies most across settings and thus best represents the changes in overall conventional commitment. Table \ref{tab:operational} shows that the deterministic solution is severely vulnerable to uncertainty for both HM and OBZ designs. Under the deterministic solution, the day-ahead cost is minimised  without regard for worst-case realisations. Accordingly, no reserves are procured and no additional generators are committed because the dispatch assumes perfect foresight. When the worst-case scenario is realised, the 
deterministic solution incurs substantial real-time redispatch costs. Under HM, corrective load shedding begins at $\Gamma_3$. Under OBZ, the deterministic solution delays the onset of load shedding to $\Gamma_4$ because OBZ's day-ahead dispatch, constrained by offshore zone boundaries, is inherently better positioned for the nodal network even without explicit robustness. The robust solution avoids or substantially reduces load shedding by procuring reserves and committing additional gas capacity, where OBZ's robust solution achieves zero load shedding under $\Gamma_3$ and limits it to 638~MW under $\Gamma_5$, while 
HM's robust solution reduces shedding but cannot eliminate it 
beyond $\Gamma_2$.

\begin{table*}[htbp]
\centering
\caption{Results for deterministic and robust solutions}
\label{tab:operational}
\footnotesize
\setlength{\tabcolsep}{3.5pt}
\begin{tabular}{@{}ll rr rr rr rr@{}}
\toprule
& & \multicolumn{2}{c}{\text{Reserve (MW)}} 
& \multicolumn{2}{c}{\text{Committed Gas Cap (MW) }}
& \multicolumn{2}{c}{\text{Real-Time Load Shed (MW)}}
& \multicolumn{2}{c}{\text{Total Inertia (MWs)}} \\
\cmidrule(lr){3-4} \cmidrule(lr){5-6} \cmidrule(lr){7-8} \cmidrule(lr){9-10}
Design & Level & Det & Robust & Det & Robust & Det & Robust & Det & Robust \\
\midrule
\multirow{5}{*}{HM}
& $\Gamma$1 & 0 & 0     & 47,849    & 49,020    & 0 & 0  & 1,632 & 1,642 \\
& $\Gamma$2 & 0 & 39,172    & 47,849    & 76,748     & 0 & 0  & 1,632 & 1,922 \\
& $\Gamma$3 & 0 & 44,893    & 47,849    & 92,588   & 3,260 & 509       & 1,632 & 2,012 \\
& $\Gamma$4 & 0 & 49,707    & 47,849    & 97,784    & 3,705 & 996  & 1,632 & 2,057 \\
& $\Gamma$5 & 0 & 53,633    & 47,849   & 100,034  & 5,213 & 2,532     & 1,632 & 3,239 \\
\midrule
\multirow{5}{*}{OBZ}
& $\Gamma$1 & 0 & 5,500        & 48,004    & 53,236   & 0 & 0  & 1,557 & 1,687 \\
& $\Gamma$2 & 0 & 32,580    &  48,004   & 80,001     & 0 & 0 & 1,557 & 1,947 \\
& $\Gamma$3 & 0 & 48,584    & 48,004   & 92,646    & 0 & 0  & 1,557 & 2,037 \\
& $\Gamma$4 & 0 & 53,536    & 48,004 & 95,887   & 987 & 464     & 1,577 & 2,087 \\
& $\Gamma$5 & 0 & 57,596    & 48,004  & 99,832   & 4,595 & 638  & 1,577 & 3,229\\
\bottomrule
\end{tabular}
\end{table*}

\begin{figure*}[htbp]
\centering
\begin{tikzpicture}

\begin{axis}[
    name=mainax,
    ybar stacked,
    width=0.95\textwidth,
    height=7cm,
    bar width=12pt,
    ylabel={Cost [kEUR]},
    xlabel={Uncertainty Level},
    xtick={1,2,3,4,5},
    xticklabels={$\Gamma$1,$\Gamma$2,$\Gamma$3,$\Gamma$4,$\Gamma$5},
    xmin=0.5, xmax=5.5,
    ymin=0,
    ymax=22000,
    scaled y ticks=base 10:0,                      
    ymajorgrids=true,
    grid style={dashed,gray!30},
    bar shift=-10pt,
    legend image post style={xscale=1, yscale=1},
    legend style={
        at={(0.5,-0.23)},
        anchor=north,
        font=\scriptsize,
        legend columns=3
    },
]

\addplot+[fill=orange!55, draw=orange!80, forget plot] coordinates {
    (1,6164) (2,6164) (3,6164) (4,6164) (5,6164)
};

\addplot+[fill=red!65, draw=red!85, forget plot] coordinates {
    (1,1903) (2,6657) (3,11511) (4,12728) (5,14512)
};

\addlegendimage{area legend, fill=orange!55, draw=orange!80}
\addlegendentry{HM DA Cost (Det.)}

\addlegendimage{area legend, fill=red!65, draw=red!85}
\addlegendentry{HM Nodal Redisp.\ (Det.)}

\addlegendimage{area legend, fill=cyan!45, draw=cyan!70!black}
\addlegendentry{OBZ DA Cost (Det.)}

\addlegendimage{area legend, fill=blue!55, draw=blue!75!black}
\addlegendentry{OBZ Nodal Redisp.\ (Det.)}

\addlegendimage{only marks, mark=diamond*, mark size=3pt, black}
\addlegendentry{HM Total (Rob.)}

\addlegendimage{only marks, mark=square*, mark size=3pt, black}
\addlegendentry{OBZ Total (Rob.)}

\end{axis}

\begin{axis}[
    ybar stacked,
    width=0.95\textwidth,
    height=7cm,
    bar width=12pt,
    xmin=0.5, xmax=5.5,
    ymin=0,
    ymax=22000,
    axis x line=none,
    axis y line=none,
    xtick=\empty,
    ytick=\empty,
    bar shift=+10pt,
]

\addplot+[fill=cyan!45, draw=cyan!70!black, forget plot] coordinates {
    (1,6183) (2,6183) (3,6183) (4,6183) (5,6183)
};

\addplot+[fill=blue!55, draw=blue!75!black, forget plot] coordinates {
    (1,2055) (2,6338) (3,7970) (4,9595) (5,13204)
};

\end{axis}

\begin{axis}[
    width=0.95\textwidth,
    height=7cm,
    xmin=0.5, xmax=5.5,
    ymin=0,
    ymax=22000,
    axis x line=none,
    axis y line=none,
    xtick=\empty,
    ytick=\empty,
    clip=false,
]

\addplot[
    only marks,
    mark=diamond*,
    mark size=3pt,
    black,
    xshift=-10pt,
    forget plot
] coordinates {
    (1,7956) (2,9354) (3,10143) (4,10956) (5,12101)
};

\addplot[
    only marks,
    mark=square*,
    mark size=3pt,
    black,
    xshift=10pt,
    forget plot
] coordinates {
    (1,7621) (2,8766) (3,8692) (4,9384) (5,11045)
};

\end{axis}

\end{tikzpicture}
\caption{Cost decomposition for HM and OBZ designs}
\label{fig:cost_decomp_hm_obz}
\end{figure*}
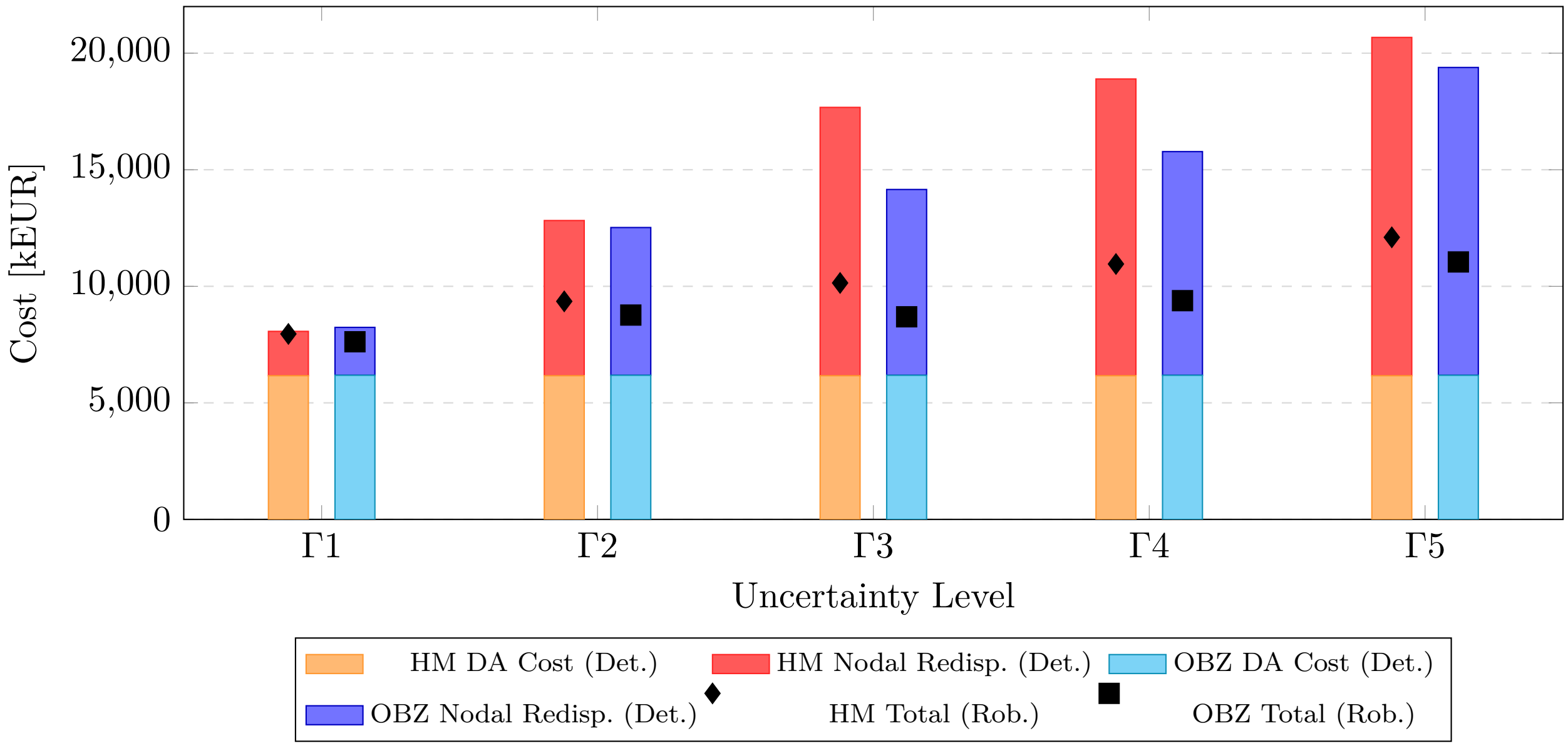

The costs resulting from day-ahead and redispatch actions under the deterministic and robust solutions for HM and OBZ designs are depicted in Figure \ref{fig:cost_decomp_hm_obz}. For each uncertainty level, the left stacked bar shows the deterministic HM cost decomposition, and the right stacked bar shows the
deterministic OBZ decomposition. Markers indicate the robust total cost for each design. When the total costs of deterministic and robust solutions are compared, it is seen that the value of robustness, which is the difference between the costs of deterministic and robust solutions, grows rapidly with the uncertainty level. 
Beyond $\Gamma_1$, the deterministic total cost exceeds the robust total cost by 37--76\%, with 2--5 times higher redispatch costs, for both designs. When results obtained for HM and OBZ designs are compared, the robust OBZ total cost is consistently lower than the robust HM total cost across all uncertainty levels. Furthermore, the deterministic OBZ total cost is also lower than the deterministic HM total cost from $\Gamma_2$ onwards, because HM's nodal redispatch cost grows faster due to its lack of offshore-targeted reserves. This demonstrates that OBZ benefits from robust optimisation not only through lower robust costs but also through inherently better-positioned day-ahead decisions that reduce vulnerability even under deterministic planning.

\subsubsection{Comparison of HM vs OBZ designs}
To compare HM and OBZ designs, we first analyse the system costs obtained for both designs under different uncertainty settings. An average winter day is investigated, i.e., solar capacity factor of 1-6\% with deviation of 50\%, offshore wind capacity factor of 50-55\% with deviation of 25\%, onshore wind capacity factor 18-25\% with deviation of 20\%, and load factor of 90\% with deviation of 15\% are considered. Table~\ref{tab:main_results} presents the full cost comparison, where the first column (\textit{Level}) presents the uncertainty level considered, the second and third columns provide the associated day-ahead (\textit{DA}), nodal redispatch (\textit{Nodal}), and total (\textit{Total}) costs for HM and OBZ designs, respectively, and the last column (\textit{OBZ Advantage}) presents the percentage reduction in the total costs when OBZ design is employed instead of HM design. 
As seen from the table, OBZ's DA cost is always slightly higher than HM's due to the additional inter-zonal constraints from offshore zone boundaries. However, this increase is far outweighed by the reduction in nodal redispatch cost and thus OBZ consistently reduces the true total cost. This result extends the findings of \cite{kenis2023off}, which show that in a deterministic setting the choice of bidding zone design is merely a welfare transfer with a constant total of DA and redispatch costs. In our robust optimisation framework, the total cost is  not constant across designs because commitment decisions and frequency security constraints can result in fundamentally different feasibility regions for the nodal redispatch.

\begin{table*}[htbp]
\centering
\caption{Cost comparison between HM and OBZ (kEUR)}
\label{tab:main_results}
\renewcommand{\arraystretch}{1.1}
\begin{tabular}{l rrr rrr r}
\hline\hline
& \multicolumn{3}{c}{\text{Home Market (HM)}} & \multicolumn{3}{c}{\text{Offshore Bidding Zone (OBZ)}} &  \\
\cmidrule(lr){2-4} \cmidrule(lr){5-7} 
{Level} & {DA} & {Nodal} & \text{Total} & \text{DA} & \text{Nodal} & \text{Total}  & \text{OBZ Advantage (\%)} \\
\hline
$\Gamma_0$      & 6{,}164  & 1{,}756 & 7{,}920 & 6{,}183  & 1{,}536  & 7{,}719   &  2.5 \\
$\Gamma_1$          & 6{,}223  &  1{,}732 & 7{,}956 & 6{,}251  & 1{,}370  & 7{,}621   & 4.2 \\
$\Gamma_2$           & 6{,}545  &  2{,}809 & 9{,}354 & 6{,}607  & 2{,}159  & 8{,}766   & 6.3 \\
$\Gamma_3$       & 6{,}709  &  4{,}338 & 11{,}046 & 6{,}779  & 1{,}913  & 8{,}692   & 21.3 \\
$\Gamma_4$      & 6{,}775 &  4{,}099 & 10{,}874 & 6{,}847 & 2{,}537  & 9{,}384    &  13.7 \\
$\Gamma_5$    & 6{,}831 &  5{,}873 & 12{,}703 & 8{,}162 & 2{,}884  & 11{,}045   &  13.1 \\
\hline\hline
\end{tabular}
\end{table*}


The cost advantage of OBZ, as a function of increasing uncertainty, shows a non-monotonic pattern because it increases from $\Gamma$0 to $\Gamma$3 and then decreases as demand uncertainty grows further. The reason for this pattern is 
related to the nature of the dominant risk. Under low to moderate uncertainty ($\Gamma_1$--$\Gamma_3$), the dominant risk is localised, e.g., a specific converter trips or regional offshore wind deviates from the forecast. OBZ's offshore zone boundaries make these localised risks visible to the market, enabling targeted reserve procurement and efficient DC flow scheduling. In particular, the sharp increase in the OBZ advantage from $\Gamma_2$ to $\Gamma_3$ occurs because adding demand uncertainty amplifies the impact of the converter contingency, i.e., HM's nodal redispatch cost nearly doubles as it lacks offshore-targeted reserves, while OBZ's nodal cost actually decreases because the additional reserves procured against demand uncertainty also cover the converter contingency. Notably, OBZ's total cost slightly decreases from $\Gamma_2$ to $\Gamma_3$, even though uncertainty increases. This decrease originates solely in the nodal redispatch stage, which is a post-processing evaluation, whereas the zonal recourse cost increases monotonically as expected from 1{,}280 to 1{,}811 kEUR. The nodal cost decreases because the additional reserves procured against demand 
uncertainty under $\Gamma_3$ substitute for costly emergency gas starts in the nodal redispatch.
Under high uncertainty ($\Gamma_4$--$\Gamma_5$), the dominant 
risk becomes system-wide because all RES generators can simultaneously deviate while demand surges across many load buses. Both designs commit nearly all available conventional generation 
and procure similar reserve volumes, reducing the relative benefit of OBZ's zone-specific visibility. Under $\Gamma_5$, OBZ's day-ahead cost rises sharply due to the preventive load shedding in the day-ahead market, yet OBZ still saves 13.1\% in total cost because HM defers the issue to the redispatch stage, where significantly more involuntary corrective load shedding occurs.

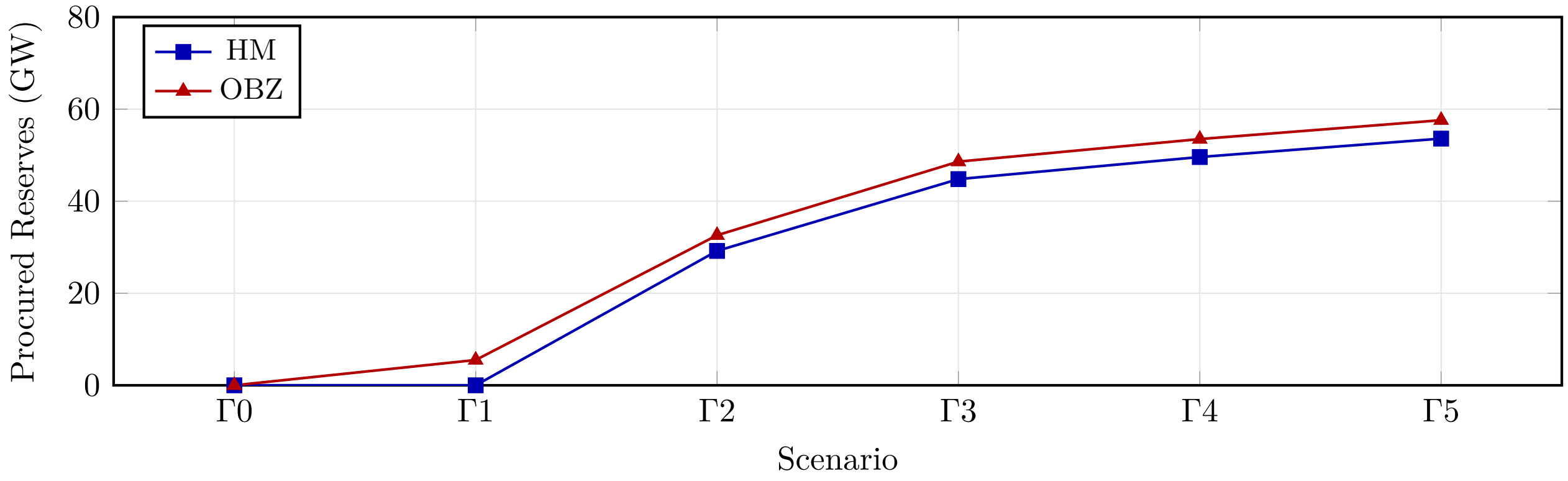
\begin{figure}[htbp]
\centering
\begin{tikzpicture}
\begin{axis}[
    width=\columnwidth,
    height=5.5cm,
    ylabel={Procured Reserves (GW)},
    xlabel={Scenario},
    symbolic x coords={$\Gamma$0, $\Gamma$1,$\Gamma$2,$\Gamma$3,$\Gamma$4,$\Gamma$5},
    xtick=data,
    ymin=0,
    ymax=80,
    legend style={at={(0.02,0.98)}, anchor=north west, font=\footnotesize},
    grid=major,
    grid style={gray!20},
    xticklabel style={font=\small},
    yticklabel style={font=\small},
    ylabel style={font=\small},
    xlabel style={font=\small},
    thick,
]
\addplot[mark=square*, blue!70!black, thick] coordinates {
    ($\Gamma$0, 0) ($\Gamma$1, 0) ($\Gamma$2, 29.2) ($\Gamma$3, 44.8) ($\Gamma$4, 49.6) ($\Gamma$5, 53.6)
};
\addplot[mark=triangle*, red!70!black, thick] coordinates {
    ($\Gamma$0, 0) ($\Gamma$1, 5.5) ($\Gamma$2, 32.6) ($\Gamma$3, 48.6) ($\Gamma$4, 53.5) ($\Gamma$5, 57.6)
};
\legend{HM, OBZ}
\end{axis}
\end{tikzpicture}
\caption{Procured reserves (GW) as a function of uncertainty level}
\label{fig:reserves}
\end{figure}

Fig.~\ref{fig:reserves} shows the procured reserves as uncertainty increases. Both designs respond to increasing uncertainty by committing more generators and procuring more reserves. Under $\Gamma$1, when the contingency is introduced to the planning, the reserve procurement asymmetry becomes the highest. This difference arises because OBZ's offshore zone boundaries make the converter contingency cost visible in the day-ahead objective function. While the market under OBZ sees the actual risk of losing lines connected to a converter and thus procures reserves to cover this risk, under HM, the same converter trip affects some or all lines that are all classified as intra-zonal, so the market has no incentive to hedge against the loss. As uncertainty increases from $\Gamma_2$ to $\Gamma_5$, the gap narrows as both designs commit more conventional generation to cover system-wide risks, but OBZ consistently procures more reserves than HM across all levels. This persistent gap reflects OBZ's continued ability to target reserves specifically for offshore contingencies, even when system-wide uncertainty dominates the reserve requirement.

We also investigate the price signals obtained under HM and OBZ market designs. Table~\ref{tab:prices} compares two measures of price signal quality, such as the price spread and the congestion rent, across designs and uncertainty levels. The price spread (EUR/MWh) is defined as the difference between the highest and lowest zonal clearing price across the system and measures the degree to which the market produces differentiated locational signals reflecting transmission congestion. The congestion rent (EUR/h) is the total revenue collected from price differences across HVDC links, computed as $\sum\limits_{l \in L} f_l (\lambda_{z_r} - \lambda_{z_s})$, where $f_l$ is the flow on link~$l$ and $\lambda_{z_s}$, $\lambda_{z_r}$ are the clearing prices in the sending and receiving zones. Zonal prices are extracted by fixing the binary commitment variables at their optimal values, relaxing the master problem to a linear program, and retrieving the dual variables of the zonal power balance constraints.

\begin{table}[htbp]
\centering
\caption{Price signal characteristics}
\label{tab:prices}
\renewcommand{\arraystretch}{1.15}
\begin{tabular}{l r r r r}
\hline\hline
& \multicolumn{2}{c}{\text{Price Spread (EUR/MWh)}} & \multicolumn{2}{c}{\text{Congestion Rent (EUR/h)}} \\
\cmidrule(lr){2-3} \cmidrule(lr){4-5}
\text{Level} & \text{HM} & \text{OBZ} & \text{HM} & \text{OBZ} \\
\hline
$\Gamma$0 & 48.18 & 50.17 & 17{,}668 & 72{,}126 \\
$\Gamma$1 & 0.00 & 3.52 & 0 & 6{,}201 \\
$\Gamma$2 & 4.66 & 4.90 & 0 & 7{,}587 \\
$\Gamma$3 & 4.74 & 4.97 & 1{,}767 & 7{,}583 \\
$\Gamma$4 & 4.82 & 4.98 & 1{,}767 & 7{,}597 \\
$\Gamma$5 & 4.82 & 4.99 & 1{,}767 & 4{,}795 \\
\hline\hline
\end{tabular}
\end{table}

OBZ produces higher congestion rent than HM in every setting, typically by a factor of 3--4. These differences arise because OBZ's five offshore zones create additional zone boundaries along the HVDC lines connecting offshore hubs to shore. Under HM, these lines are intra-zonal and therefore carry no price differential regardless of how congested they are. Under OBZ, the same lines become inter-zonal, and any congestion produces a price spread that generates congestion rent reflecting the true scarcity of transmission capacity.
The most revealing case happens when the uncertainty level $\Gamma_1$. HM provides uniform prices across all zones, i.e., spread and congestion rent are 0. This occurs because the robust solution, anticipating a converter trip, pushes DC flows to maximise frequency response headroom on the surviving lines. Since all DC lines are intra-zonal under HM, this flow redistribution does not create any inter-zonal congestion, and the market sees no bottleneck and clears at a single system-wide price. OBZ, by contrast, maintains a price spread of 3.52~EUR/MWh and generates a positive congestion rent under the same contingency budget, because the offshore zone boundaries preserve locational signals even when DC flows are optimised for frequency security.
As uncertainty increases from $\Gamma_2$ to $\Gamma_5$, both designs recover some price differentiation because the interaction of multiple uncertainty sources creates congestion on the AC inter-zonal corridors. Under HM with $\Gamma_2$, the price spread reaches 4.66~EUR/MWh, yet DC congestion rent remains zero because the price differences arise entirely from AC corridor congestion, while the offshore DC lines generate no rent regardless of their loading. From $\Gamma_3$ onwards, HM generates a modest congestion, but this remains 
far below OBZ's because OBZ captures rent on both the offshore HVDC lines and the onshore AC corridors, while HM can only price congestion on the latter.

These price results have direct implications for offshore infrastructure financing. When electricity flows through a line from a low-price area to a high-price area, the resulting congestion rent can be used to cover the cost of building and maintaining the line. Under OBZ, the offshore hub and the onshore grid are in different bidding zones, so whenever the line is heavily used, a price difference appears and the line earns revenue. Under HM, however, the offshore hub and the onshore grid are in the same bidding zone. Because both ends of the line see the same market price, the line earns zero revenue from congestion, no matter how much power flows through it. This means that under HM, there is no market-based way to recover the cost of offshore lines. However, OBZ avoids this problem by allowing the market itself to generate revenue for line owners in proportion to how much the line is actually needed by the system. Accordingly, the revenue obtained from congestion can be reinvested to expand and improve offshore infrastructure, as proposed by the European Grids Package (\cite{congestion}).

We lastly conduct a set of computational experiments considering the results 
obtained over a full day, i.e., 24 hours, under only converter contingency ($\Gamma$1). Two representative days are considered: 
a winter (December) day and a summer (June) day. Across the hours of each day, capacity factors 
for solar and wind generation, as well as load factors, are adjusted accordingly. For example, the average capacity factor of solar is 3\% and 19\%, the average capacity factor of offshore wind is 39\% and 20\%, the average capacity factor of onshore wind is 30\% and 14\%, and the average load factor is 92\% and 66\% for December and June, respectively.
Note that since the market clearing problem studied in this paper is separable 
across hours, each hour is solved independently. Additionally, to isolate the 
effect of zone design from that of contingency selection, we consider a common 
worst-case contingency scenario identified by the OBZ oracle at each hour and provide it to HM design as given. This removes any advantage that one design 
might gain from facing a less severe contingency, ensuring that all cost 
differences are attributable solely to how each zonal market perceives and 
responds to the same physical event.

 
\begin{figure}[htbp]
\centering
\begin{tikzpicture}
\begin{axis}[
    ybar,
    bar width=7pt,
    width=\columnwidth,
    height=5.5cm,
    ylabel={OBZ Advantage (kEUR)},
    xlabel={Hour of Day},
    xmin=0.5, xmax=24.5,
    xtick={2,4,6,8,10,12,14,16,18,20,22,24},
    xticklabel style={font=\scriptsize},
    yticklabel style={font=\scriptsize},
    ylabel style={font=\small},
    xlabel style={font=\small},
    ymin=-0, ymax=1100,
    ytick scale label code/.code={$\cdot 10^{3}$},   
    grid=major, grid style={gray!12},
    extra y ticks={0},
    extra y tick style={grid=major, grid style={black, thick}},
    legend style={at={(0.02,0.98)}, anchor=north west, font=\tiny, fill=white, fill opacity=0.9, draw=gray!40},
]
\addplot[fill=green!50!black, fill opacity=0.6, draw=green!70!black] coordinates {
    (1,394.4) (2,389.9) (3,362.3) (4,384) (5,357) (6,341.9)
    (7,14.5) (8,424.1) (9,1037.1) (10,231.1) (11,547.7)
    (12,408.9) (13,382.2) (14,448.4)
    (15,501.8) (16,37.9) (17,467.3) (18,674.6) (19,647.4)
    (20,662.6) (21,1003.4) (22,714.3) (23,589.7) (24,191.5)
};
\end{axis}
\begin{axis}[
    width=\columnwidth, height=5.5cm,
    xmin=0.5, xmax=24.5, ymin=0, ymax=66,
    axis y line=right, axis x line=none,
    ylabel={Offshore Wind CF (\%)},
    ylabel style={font=\small, gray!60!black},
    yticklabel style={font=\scriptsize, gray!60!black},
    ytick={0,30,60},
]
\addplot[gray!50!black, thick, dashed, mark=none] coordinates {
    (1,34) (2,33) (3,31) (4,33) (5,30) (6,28)
    (7,29) (8,32) (9,34) (10,36) (11,35) (12,33)
    (13,31) (14,34) (15,38) (16,42) (17,48) (18,54)
    (19,56) (20,53) (21,59) (22,51) (23,47) (24,40)
};
\end{axis}
\end{tikzpicture}
\caption{Hourly OBZ cost advantage for a day in December}
\label{fig:paired_dec}
\end{figure}
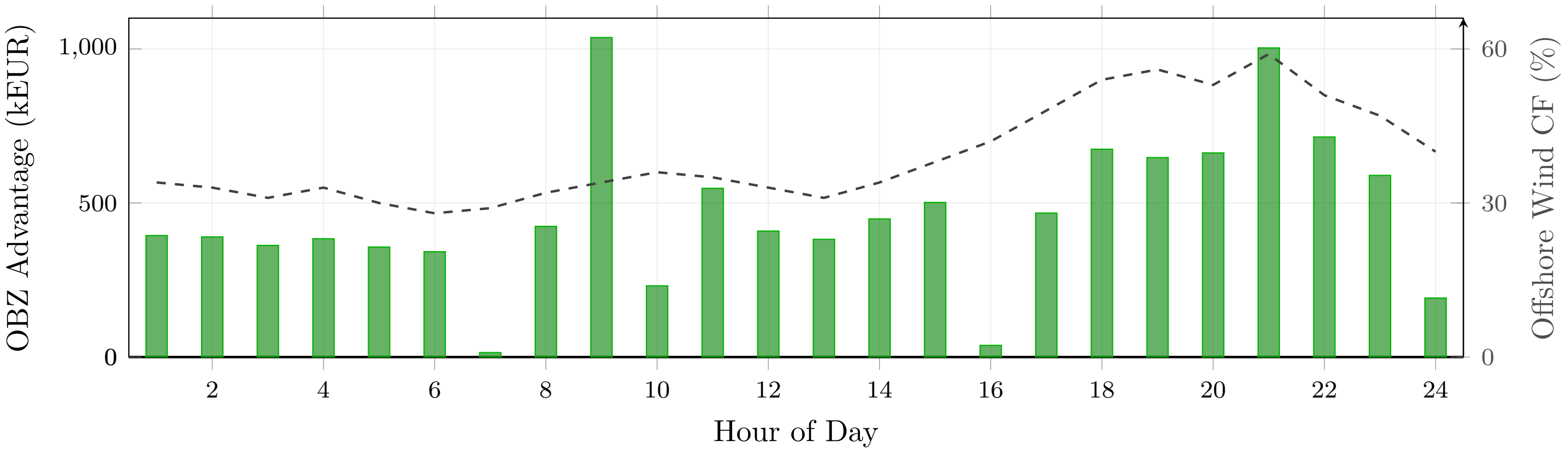


\begin{figure}[htbp]
\centering
\begin{tikzpicture}
\begin{axis}[
    ybar,
    bar width=7pt,
    width=\columnwidth,
    height=5.5cm,
    ylabel={OBZ Advantage (kEUR)},
    xlabel={Hour of Day},
    xmin=0.5, xmax=24.5,
    xtick={2,4,6,8,10,12,14,16,18,20,22,24},
    ytick={200,400},
    xticklabel style={font=\scriptsize},
    yticklabel style={font=\scriptsize},
    ylabel style={font=\small},
    xlabel style={font=\small},
    ymin=-0, ymax=480,
    ytick scale label code/.code={$\cdot 10^{3}$},   
    grid=major, grid style={gray!12},
    extra y ticks={0},
    extra y tick style={grid=major, grid style={black, thick}},
    legend style={at={(0.02,0.98)}, anchor=north west, font=\tiny, fill=white, fill opacity=0.9, draw=gray!40},
]
\addplot[fill=green!50!black, fill opacity=0.6, draw=green!70!black] coordinates {
    (1,51.9) (2,209.2) (3,125.2) (4,87.1) (5,73.7) (6,112.6)
    (7,56.4) (8,164.1) (9,70.3) (10,57) (11,178.4)
    (12,191.7) (13,194.4) (14,294)
    (15,144.7) (16,388.3) (17,131.2) (18,170.7) (19,176.5)
    (20,76.9) (21,205.1) (22,130.5) (23,261.3) (24,161.5)
};
\end{axis}
\begin{axis}[
    width=\columnwidth, height=5.5cm,
    xmin=0.5, xmax=24.5, ymin=0, ymax=72,
    axis y line=right, axis x line=none,
    ylabel={Offshore Wind CF (\%)},
    ylabel style={font=\small, gray!60!black},
    yticklabel style={font=\scriptsize, gray!60!black},
    ytick={0,30,60},
]
\addplot[gray!50!black, thick, dashed, mark=none] coordinates {
    (1,24) (2,25) (3,26) (4,23) (5,22) (6,20)
    (7,19) (8,18) (9,16) (10,15) (11,14) (12,14)
    (13,15) (14,16) (15,17) (16,18) (17,19) (18,20)
    (19,21) (20,22) (21,25) (22,26) (23,25) (24,23)
};
\end{axis}
\end{tikzpicture}
\caption{Hourly OBZ cost advantage for a day in June}
\label{fig:paired_jun}
\end{figure}
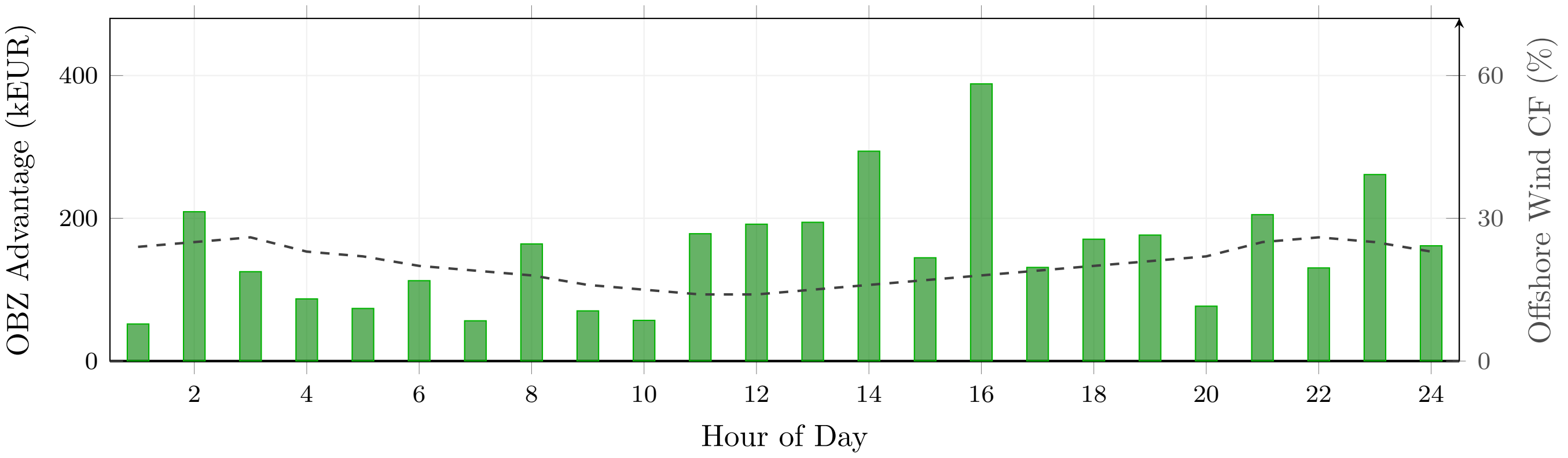
 
Figures \ref{fig:paired_dec} and \ref{fig:paired_jun} show the hourly OBZ savings as bars and offshore wind availability as a dashed line for the winter and summer days, respectively.  Table \ref{tab:paired_seasonal} presents the aggregated results, over 24 hours, including day-ahead costs, nodal redispatch costs, 
real-time load shedding, and reserve activation for both designs. As seen from Figure \ref{fig:paired_dec}, the savings provided by OBZ are becoming considerably higher for hours when the offshore wind generation is high. The most striking result is that HM procures
zero reserves across all 24 hours as seen from Table~\ref{tab:paired_seasonal}. This occurs because the OBZ-identified contingency trips an offshore converter and its associated lines, all of which are intra-zonal under HM.  The HM zonal market cannot distinguish this
contingency from normal operation, i.e., the tripped lines connect the offshore hub to shore within the same bidding zone, so no inter-zonal flow is affected, and no price signal triggers reserve procurement.
Consequently, the entire burden of the converter trip falls on the nodal redispatch stage as involuntary load shedding. Under OBZ, the same
lines are inter-zonal, making the contingency visible to the day-ahead
market. The OBZ responds by procuring more reserves daily, avoiding extra load shedding that would otherwise incur more costs.
 
\begin{table}[htbp]
\centering
\caption{Aggregated results under OBZ-identified contingency for each hour}
\label{tab:paired_seasonal}
\renewcommand{\arraystretch}{1.1}
\begin{tabular}{l rr rr}
\hline\hline
& \multicolumn{2}{c}{\text{December}} & \multicolumn{2}{c}{\text{June}} \\
\cmidrule(lr){2-3} \cmidrule(lr){4-5}
\textbf{} & \text{HM} & \text{OBZ} & \text{HM} & \text{OBZ} \\
\hline
RT load shedding (MWh)   & 27{,}501 & 25{,}626 & 0 & 0 \\
Reserve activation (MWh) & 0         & 55{,}800  & 0        & 48{,}200 \\
DA cost (kEUR)             & 181{,}881 & 183{,}064 & 79{,}584  &  80{,}856\\
Nodal redispatch  cost (kEUR)  & 69{,}718 & 57{,}321 & 23{,}197 &  18{,}212\\
Total cost (kEUR)& 251{,}599 & 240{,}385 & 102{,}781 & 99{,}068 \\
\hline
OBZ advantage  (\%)          & \multicolumn{2}{c}{\text{4.5}} & \multicolumn{2}{c}{\text{3.6}} \\
\hline\hline
\end{tabular}
\end{table}

The results for a summer day are presented in Figure~\ref{fig:paired_jun}. Both HM and OBZ achieve zero load shedding throughout the day. The OBZ advantage is smaller 
than in winter but remains consistently positive. 
In June, solar PV provides up to a 60\% capacity factor at midday while offshore wind drops to 14--15\%. Since solar generation is located onshore and does not flow
through the HVDC lines connecting offshore hubs to shore, the converter
contingency produces less line congestion, reducing the magnitude of 
OBZ's hourly savings during midday hours. The largest 
savings occur at hours when either offshore wind is 
relatively high (hours~1--3, 21--24) or peak loads create 
high line utilisation (hours~12--16). The key operational 
discrepancy persists across all hours for both designs, i.e., HM procures zero 
reserves and relies entirely on emergency gas starts 
for redispatch, while OBZ activates pre-procured reserves and avoids emergency starts entirely. 
This confirms that although the magnitude of OBZ advantage varies with seasons in absolute terms, the advantage is structural and persists.

These daily results yield a clear policy implication which is that OBZ's value is directly
proportional to the ratio of offshore wind output to total HVDC line
capacity. When this ratio approaches or exceeds unity, e.g., winter
evening, lines saturate, and OBZ's congestion visibility becomes
critical. When the ratio is low, e.g., summer midday, onshore solar
dominates the generation mix, lines are lightly loaded, and OBZ's 
advantage becomes smaller.

\section{Conclusion} \label{sec:conclusion}

The rapid expansion of offshore wind in the North Sea will create a meshed HVDC transmission network  whose converter stations introduce large single-point-of-failure  contingencies. Current European market practice clears the day-ahead  market without co-optimizing energy dispatch, reserve procurement, and frequency security, and manages converter contingencies through separate deterministic N-1 screening. At the same time, the question of whether offshore wind farms should be assigned to their nearest onshore bidding zone (HM) or form dedicated offshore bidding zones (OBZ) remains an open policy debate with direct consequences 
for dispatch efficiency, reserve procurement, price formation, and infrastructure financing.
This paper addresses both questions within a unified framework by formulating the day-ahead market clearing as a two-stage adaptive robust optimisation problem. The first stage determines unit commitment, energy dispatch, and reserve procurement under flow-based zonal coupling, and the second stage represents worst-case balancing actions under simultaneous uncertainty in renewable generation, load, and HVDC converter outages. Frequency security is enforced explicitly through inertia and fast frequency response constraints at the synchronous-area level, with surviving HVDC lines providing emergency power support within their available headroom. The post-fault active-power imbalance entering these constraints is endogenous to 
the day-ahead schedule, meaning that the severity of a converter trip depends on how much power was flowing through the affected lines at the time of failure. The resulting min--max--min problem is solved by a column-and-constraint generation algorithm in which the bilinear oracle objective is linearised exactly using McCormick envelopes. The framework is applied to an eight-country North Sea test system comprising three synchronous areas, sixteen onshore bidding zones, and five offshore wind farm hubs connected through a meshed HVDC grid.

The computational experiments yield several findings. 
Firstly, robust solutions substantially outperform deterministic solutions. When the deterministic day-ahead dispatch is exposed to worst-case uncertainty realisations, the resulting redispatch cost is at least twice that of the robust solution in every scenario beyond the pure contingency case. 
Secondly, OBZ reduces total system cost by 2--21\% compared to HM across all tested uncertainty levels. The savings are driven entirely by cheaper nodal redispatch, 
as OBZ's day-ahead cost is consistently slightly higher due to the additional inter-zonal constraints from offshore zone boundaries.  
The OBZ advantage peaks under $\Gamma_3$, where low demand uncertainty is combined with converter contingency and RES variability. Under higher system-wide stress, such as $\Gamma_4$ and $\Gamma_5$, the advantage remains around 13\%, as OBZ continues to provide more targeted reserve procurement.
Thirdly, OBZ generates significantly higher congestion rent than HM in every scenario, providing a market-based revenue stream for offshore line
owners that HM cannot produce by construction, since the lines 
connecting offshore hubs to shore are intra-zonal and thus 
invisible to the market. 
Lastly, the paired contingency experiments, 
in which both designs face the identical worst-case event, reveal a 
critical asymmetry. In the winter scenario, HM procures zero 
reserves and relies entirely on emergency gas turbine starts at three times the marginal cost, while OBZ procures reserves and uses zero emergency gas, with both designs ending with similar corrective load shedding. In the summer scenario, OBZ 
saves 3.6\% with both designs achieving zero load shedding, 
while in winter OBZ saves 4.5\%. The hourly and seasonal variations in savings confirm that OBZ's value is directly proportional to 
the ratio of offshore wind output to HVDC line capacity. As the North Sea offshore wind capacity expands, this ratio will increase, making the OBZ design more advantageous for efficient market clearing and frequency security.

Several promising future research directions emerge from this work. First, integrating energy storage flexibility and extending the framework to a multi-period formulation would capture the more realistic representation of electricity markets. 
Secondly, different uncertainty models can be employed because the studied polyhedral uncertainty sets may overestimate worst-case costs relative to distributional approaches. In this regard, a distributionally robust extension could provide tighter bounds, but it will cause some computational burdens that need to be studied. Another direction can be to embed the 
market-clearing model within a capacity expansion framework, where the congestion rent generated under OBZ would directly inform the economic case for new offshore line investments, linking zone design to grid infrastructure planning. 

\section*{Acknowledgements}

This work is conducted within the framework of the CROCODILE project funded by the Belgian Energy Transition Funds.

\newpage
\bibliographystyle{abbrvnat}
\bibliography{reference}
\end{document}